\documentclass{elsarticle}

\def\BibTeX{{\rm B\kern-.05em{\sc i\kern-.025em b}\kern-.08emT\kern-.1667em\lower.7ex\hbox{E}\kern-.125emX}}

\usepackage[utf8]{inputenc}
\usepackage{textcomp}

\usepackage{amssymb}

\usepackage{xcolor}
\usepackage{comment}

\usepackage{subcaption}
\usepackage{graphicx}

\usepackage{booktabs}
\usepackage{url}

\usepackage[utf8]{inputenc}
\usepackage{booktabs}
\usepackage{dcolumn,tabularx,ragged2e}
\usepackage{siunitx}
\usepackage{tablefootnote}

\usepackage{multirow,tabularx}

\usepackage{adjustbox}

\usepackage{longtable}

\newif\ifdraft
\drafttrue

\journal{Elsevier Information and Software Technology}

\begin{document}

\begin{frontmatter}

\title{Analyzing developer discussions on EU and US privacy legislation compliance in GitHub repositories}

\author{Georgia M. Kapitsaki\textsuperscript{a,}\corref{cor1}}
\ead{gkapi@ucy.ac.cy}
\author{Maria Papoutsoglou\textsuperscript{a}}
\ead{mpapoutsogloy@gmail.com}
\author{Christoph Treude\textsuperscript{b}}
\ead{ctreude@gmail.com}
\author{Ioanna Theophilou\textsuperscript{a}}
\ead{theophilou.ioanna@ucy.ac.cy}

\address{\textsuperscript{a}University of Cyprus, 1, University Avenue, Aglantzia, 2109, Cyprus}
\address{\textsuperscript{b}Singapore Management University, Stamford Road, Singapore, 178903, Singapore}

\cortext[cor1]{Corresponding author}

\begin{abstract}
\emph{Context}: Privacy legislation has impacted the way software systems are developed, prompting practitioners to update their implementations. Specifically, the EU General Data Protection Regulation (GDPR) and the California Consumer Privacy Act (CCPA) have forced the community to focus on users' data privacy.\\
\emph{Objectives}: Relying on the vast amount of data on developer issues available in GitHub repositories, our aim is to gather empirical evidence on the issues developers of Open Source Software discuss to comply with privacy legislation.\\
\emph{Method}: We examined such discussions by mining and analyzing 32,820 issues from GitHub repositories. We partially analyzed the dataset automatically to identify law user rights and principles indicated, and manually analyzed a sample of 1,186 issues based on the type of concern addressed.\\
\emph{Results}: We devised 24 discussion categories placed in six clusters: user rights and consent, compliance implementation, documentation, data storing/sharing, general compliance, and contextual adaptability. Developers mainly focus on specific user rights from the legislation (right to erasure, right to opt-out, right to access), addressing other rights less frequently, while most discussions concern user consent, user rights functionality, bugs and cookie management.\\
\emph{Conclusion}: The created taxonomy can help practitioners understand which issues are discussed for law compliance, so that they ensure they address them first in their systems. In addition, the educational community can reshape curricula to better educate future engineers on the privacy law concerns raised, and the research community can identify gaps and areas for improvement to support and accelerate data privacy law compliance.\end{abstract}


\end{frontmatter}

\maketitle

\section{Introduction}

There has been a shift in data privacy legislation worldwide. Major privacy laws enacted in the last decade include the EU General Data Protection Regulation (GDPR)~\citep{voigt2017eu} and the California Consumer Privacy Act of 2018 (CCPA)~\citep{goldman2020introduction}. Other countries have followed suit by introducing laws specific to their regions, such as the Brazilian General Data Protection Law (LGPD)~\citep{de2022ensuring} and the California Privacy Rights Act (CPRA) CCPA amendment~\citep{determann2020california}. According to the United Nations Conference on Trade and Development, 79\% of countries have data protection and privacy legislation in place~\citep{unctaddata}.

Software systems must comply with privacy legislation, and development activities have been impacted by the introduction of relevant laws, which require incorporation of user rights and other principles (e.g., user right to erasure, data minimization principle)~\citep{vanezi2019gdpr}. It is crucial to investigate and understand developers' concerns and approaches to ensuring privacy law compliance in Open Source Software (OSS) repositories. Such insights can help practitioners improve their development practices, help policymakers refine regulations, and support educators in updating curricula to address these essential compliance issues effectively. 

To the best of our knowledge, there has been limited and only recent activity towards linking privacy legislation and respective repository activities. Our work complements attempts that use other sources to understand data privacy considerations in software systems, such as analyses of Reddit discussions~\citep{li2021developers,parsons2023understanding}, Q\&A sites on privacy~\citep{tahaei2020understanding}, and recent work that analyzes GitHub pull requests, commits, and issues~\citep{franke2024exploratory,kapitsaki2024gdpr,datler2023intended,sangaroonsilp2023taxonomy,hennig2026whos}. 

\begin{figure}
\centering
\includegraphics[scale=0.147]{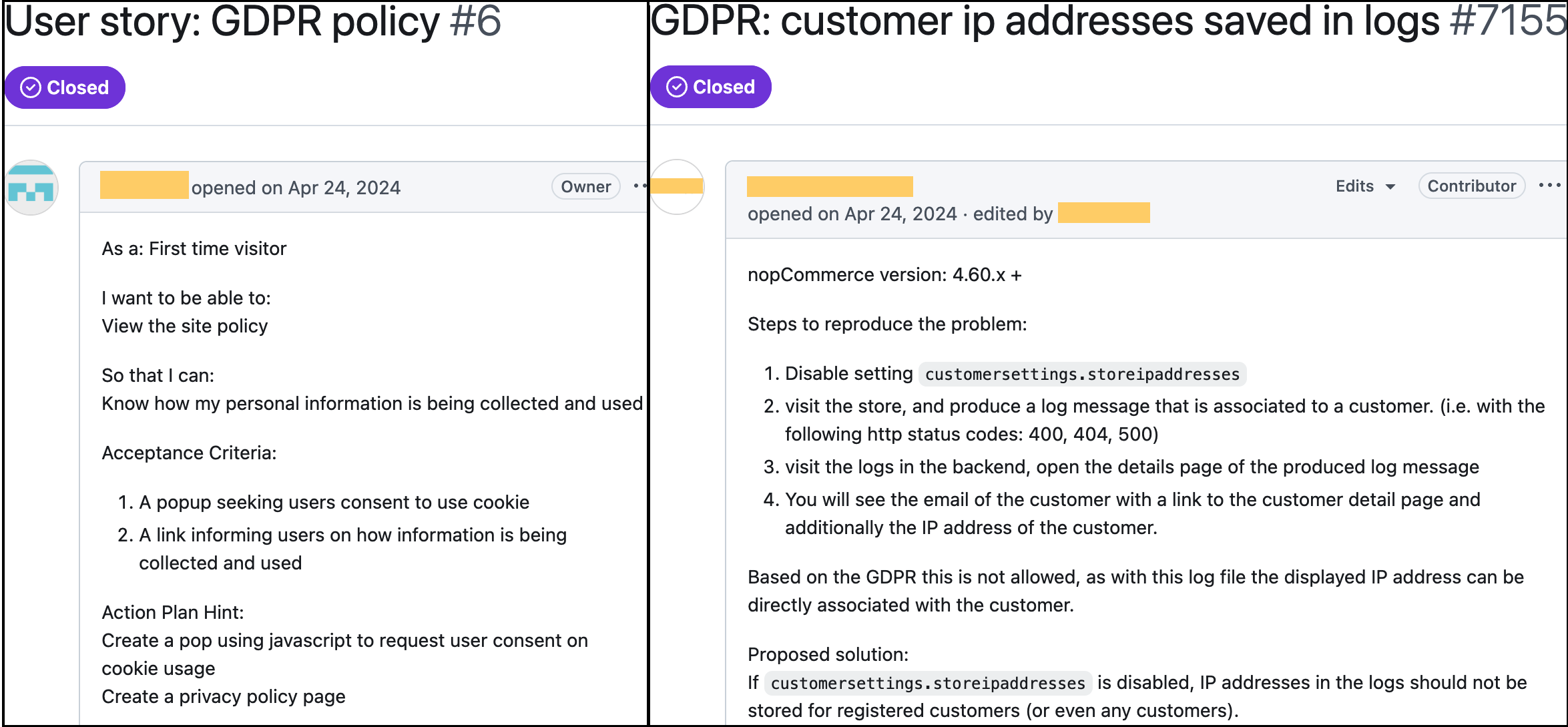}
\caption{Examples of privacy law compliance discussions in issues (usernames are hidden): user story on privacy policy, relevant to the right to information (left)~\citep{zitmall6}, and problem with storing customer IP addresses, relevant to the purpose limitation principle (right)~\citep{nopCommerce7155}.}
\label{fig:issues-examples-screenshots} 
\end{figure}

We have chosen GitHub as our data source because it hosts a vast amount of software data, providing a good reflection of the typical issues developers face when dealing with privacy legislation. Initially, we collected all available issues that mentioned privacy laws: GDPR, CCPA, CPRA, and the UK Data Protection Act 2018 (DPA). We are considering UK DPA as it sits alongside GDPR. After appropriate filtering, we were left with 32,820 issues from 13,227 repositories spanning from April 2016 (official adoption month of GDPR by the European Parliament and Council) to June 2024. We then performed automated analysis on the filtered issues, while a manual examination was conducted on a sample of 1,186 issues. We focused specifically on privacy legislation compliance, examining issues that address aspects such as user rights from the legislation (e.g., right to information) and law-specific principles (e.g., purpose limitation) because these elements represent the actionable requirements developers must implement to ensure legal compliance (two indicative example issues are depicted in~\figurename~\ref{fig:issues-examples-screenshots}).

To guide our research, we introduce the following research questions (RQs):
\begin{itemize}
\item \textbf{RQ1}. \emph{Which privacy legislation user rights and principles are present in developers' discussions?} We identified a number of user rights and principles from existing privacy laws and examined their presence in GitHub issues using both automated and manual analysis, with the manual analysis performed on a representative sample of 1,186 issues. We found that a limited number of rights are explicitly mentioned, with the right to erasure and the right to opt-out being the most popular.
\item \textbf{RQ2}. \emph{What concerns do developers discuss when dealing with privacy legislation?} To answer this RQ, we used a bottom-up thematic analysis by manually examining the issue discussions sample and, when necessary, relevant commits and source code, identifying categories of concerns developers have, such as privacy policy updates and User Interface (UI) changes. The taxonomy created from this analysis includes 24 categories placed into six clusters: user rights and consent, compliance implementation, documentation, data storing/sharing, general compliance, and contextual adaptability. We validated the taxonomy with experts and an unseen dataset and also examined differences among categories.
\end{itemize}

Our work can assist practitioners in identifying which issues are more frequent to address or discuss. It serves as an initial step towards automating the inclusion of privacy guaranties through automated recommendations that guide practitioners on handling law-relevant concerns, since we first need to know which are the main areas of concern. Although our work cannot directly indicate proven software compliance with legislation, as we relied on automated and manual analysis of issues and not source code execution, the issue discussions provide a good indication of the changes that needed to be discussed or performed, and were completed by the project contributors (only closed issues were considered).

The remainder of the text is structured as follows. Section 2 presents relevant related work, while Section 3 describes the methodological process followed. The empirical analysis for the RQs and the answers are presented in Section 4, whereas a discussion of the main findings and implications follows in Section 5. Section 6 discusses threats to validity, and finally, Section 7 concludes the paper outlining future work directions.

\section{Related work}

\subsection{Privacy topics analysis}

Earlier than GDPR's introduction,~\citep{hadar2018privacy} focused on Privacy by Design and investigated developers' views on how they approach privacy. Interviews were conducted with 27 software developers and, among other findings, it was found that developers use the vocabulary of data security to approach privacy challenges. Considering also data privacy laws, developers' discussions on Reddit were examined using qualitative analysis on 329 unique threads that mentioned different forms of personal data from the \emph{r/androiddev} forum~\citep{li2021developers}. The authors relied on four privacy laws (CCPA, CalOPPA - California Online Privacy Protection Act, COPPA - Children's Online Privacy Protection Act, and GDPR) to extract relevant terms for personal data (e.g., first name, phone number). Among the privacy topics discussed, notice/consent, choice/participation, and accountability were prominent. Another work on Reddit discussions used word frequency, topic clustering (Latent Dirichlet Allocation - LDA), and classification, analyzing 437,317 threads from \emph{r/webdev}, \emph{r/androiddev}, and \emph{r/iOSProgramming}~\citep{parsons2023understanding}. Regarding privacy laws, a significant change in topics and terms was observed due to GDPR, whereas the change due to CCPA was less significant.

LDA was applied to 1,733 privacy-related questions on Stack Overflow and then a random sample of 315 questions was manually analyzed~\citep{tahaei2020understanding}. In the qualitative analysis, drivers from laws and regulations (e.g., GDPR) were less common than other drivers, found only in 5.2\% of the questions. However, references to other regulations, such as the Health Insurance Portability and Accountability Act (HIPAA), were present in the entire dataset. In the same line of work, 119 accepted answers were used to extract privacy-related advice, with most cases referring to compliance with regulations (e.g., check if your privacy policy is compliant, inform users when requesting permissions)~\citep{tahaei2022understanding}. Thematic analysis of this sample also identified themes such as privacy policies, encryption, and access control, but there was no analysis of specific user rights and law principles. A study employing LDA and TF-IDF (Term Frequency-Inverse Document Frequency) that considered Stack Overflow, Information Security, and Software Engineering Stack Exchange sites was conducted by~\cite{diepenbrock2023analysis}. Among the topics, the \emph{Legal} topic covers discussions on compliance with laws, such as GDPR and CCPA, although the name of the laws was not used for data collection purposes.

\subsection{Repository analysis on GitHub}

Issues have been studied in earlier work to understand the overall adoption of issue trackers, the relevant categories, how they are used by the project community, and how they relate to project success~\citep{bissyande2013got}. \cite{khalajzadeh2022diverse} analyzed issue comments from 12 projects on GitHub, examining human-centric issues, and \emph{Privacy \& Security} was among the issues found. In the area of privacy, it was investigated whether positive discussions on privacy-related issues can predict privacy compliance, using GitHub and Jira issues~\citep{guber2023empirical}. The preliminary results of the work achieved around 86\% accuracy in automatically identifying privacy-related issues using supervised machine learning techniques. In this work, no law analysis was used.

A taxonomy of privacy-relevant requirements was created using GDPR, the ISO/IEC 29100 privacy framework, the Thailand Personal Data Protection Act (PDPA) and the Asia-Pacific Economic Cooperation (APEC) privacy framework~\citep{sangaroonsilp2023taxonomy}. The issue reports of Chrome and Moodle were then classified in the taxonomy, and it was found that in Moodle privacy-relevant issues were resolved faster and had fewer comments than non-privacy issues, while in Chrome they took longer to resolve and had more comments. GitHub commits relevant to privacy legislation were analyzed and various characteristics, such as differences between programming languages and terms and user rights appearing, were examined by~\cite{kapitsaki2024gdpr,kapitsaki2025evolution}.

The most relevant related work can be found in the following three studies. Datler performed an automated classification of issues as GDPR-related~\citep{datler2023intended}. A total of 541 samples were manually labeled as GDPR-relevant or not and used to train a model, resulting in 2,260 labeled issues using 16 GDPR articles and manual labeling. The articles were then grouped into: general applicability, notice and consent, data subject control rights, data security, and organizational requirements. These groups were compared on various properties, including the number of comments and reactions. A survey with developers and an analysis of pull requests were used to understand how GDPR affects Open Source Software development~\citep{franke2024first,franke2024exploratory}. Data from 56 survey participants were analyzed, with developers commenting on the impact of GDPR on data management, time and costs, design, organization, and benefits. Regarding pull requests, it was found that GDPR-related pull requests have more development activity in terms of commits, additions, deletions, and files changed, as well as review activity, 
while their sentiment remains stable over time. 652 GitHub issues were analyzed quantitatively and the relation between roles, resolutions and data protection issues were examined in a very recent work published after the completion of our work~\citep{hennig2026whos}. The authors focused on how often data protection topics are reported, what issues are reported, by whom, and how developers react. This work analyzed a smaller sample of collected issues and focused on developers' interaction, and not on the connection with user rights and principles, and specific concerns.

\subsection{Relation to previous works} 

Previous works have analyzed privacy issues using other sources of data (e.g., Reddit~\citep{li2021developers,parsons2023understanding}, Stack Overflow~\citep{tahaei2022understanding}, survey data~\citep{franke2024first,franke2024exploratory}) or have focused on different aspects of analysis (e.g., sentiment analysis and development activity of GDPR pull requests~\citep{franke2024first,franke2024exploratory}, automated classification and relation with GDPR articles~\citep{datler2023intended}, privacy-relevant requirements for software creation~\citep{sangaroonsilp2023taxonomy}, involved parties and status of GitHub issues and to a extent their content~\citep{hennig2026whos}). No previous work has focused on specific privacy compliance concerns in OSS repositories issues and relevant categorizations of developer concerns, linking issues also with data privacy legislation user rights and principles, or has examined such a large dataset of issues (issues categories are indicated in a limited way in~\citep{datler2023intended,hennig2026whos} and are compared in Section 5 of our work).

\section{Methodology}

\begin{figure}[!t]
\centering
\includegraphics[scale=0.57]{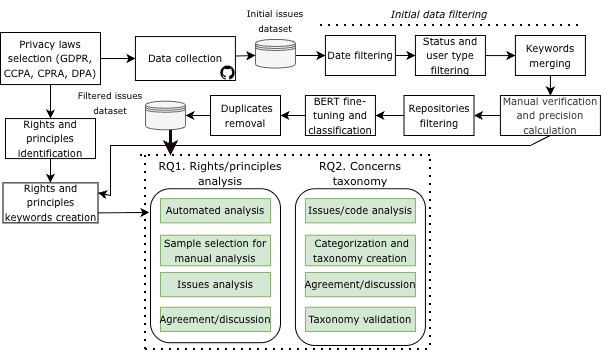}
\caption{Methodological process.}
\label{fig:methodological-process} 
\end{figure}

The aim of our work is to understand the presence of data privacy legislation in GitHub repositories and to create a taxonomy of issues discussed to comply with the legislation. The whole methodological process is depicted in~\figurename~\ref{fig:methodological-process}. Steps in white support dataset construction and pre-processing, while steps in green support the RQs.

\subsection{Privacy laws selection}

Initially, we selected the following privacy laws before collecting relevant issues available on GitHub: GDPR, CCPA, CPRA, UK DPA. We selected these laws because they are highly visible privacy regulations that have had substantial practical impact on software systems in Europe and the United States. GDPR is the privacy law applicable in EU, while UK DPA and UK GDPR sit alongside in the UK context. In the United States, there is no federal law but CCPA (and its amendment CPRA) is the closest equivalent of GDPR and operationally relevant state-level privacy regulation. Table~\ref{tab:privacy-laws} summarizes main elements of the selected laws. In addition, two of the authors have research experience with GDPR and compliance in software systems. We created the following set of keywords that needed to be present in the issue: \\
\emph{General Data Protection Regulation} and its abbreviation \emph{GDPR}, \emph{California Consumer Privacy Act} and its abbreviation \emph{CCPA}, \emph{California Privacy Rights Act} and its abbreviation \emph{CPRA}, and finally \emph{Data Protection Act}. \\
We did not consider UK GDPR separately, as it basically refers to the same provisions of the EU GDPR. We also did not use the abbreviation of Data Protection Act, DPA, due to its wider use (e.g., in Delegation of Procurement Authority, Data Processing Agreement). As our focus is on compliance with privacy legislation, we also did not include generic privacy terms such as \emph{data privacy}.

\begin{table}[h]
\centering
\begin{adjustbox}{width=0.99\textwidth}
  \begin{tabular}{llp{1.6cm}p{7.2cm}}
    \toprule
    \textbf{Law} & \textbf{Jurisdiction} & \textbf{Effective date} & \textbf{Relevance for software systems} \\
    \midrule
    GDPR & EU & May 25, 2018 & Software systems must be built with Privacy by Design, ensuring early integration of technical measures (e.g., encryption, data minimization).\\
    \midrule
    CCPA & California & January 1, 2020& Software systems must include mechanisms for opting-out of sale (of personal information) and provide automated ways to answer access requests for collected data.\\
    \midrule
    CPRA &California & July 1, 2023& Requires systems to implement stricter controls over sensitive personal information and enforces data minimization.\\
    \midrule
    UK DPA & UK & May 25, 2018 & Requires software systems to provide specific protections for law enforcement processing while maintaining data handling principles.\\
    \bottomrule
  \end{tabular}
\end{adjustbox}
\caption{Data privacy laws considered.}
\label{tab:privacy-laws} 
\end{table}

\subsection{Rights and principles identification}

As a preparatory step, we used the privacy laws considered in the study and relevant online resources~\citep{securiti} to identify the basic user rights and law principles and focus the subsequent issue analysis on these aspects~\citep{perera2019towards}. As user rights and principles from the laws are well-known, have been analyzed and made available via online resources, we did not follow a formal methodological process using text analysis of the laws for this part of the methodology~\citep{securiti}.
 
User rights are fundamental in all laws, as they provide users with control over activities on their data (shown in Table~\ref{tab:law-user-rights} with a comparison between GDPR/UK DPA and CCPA/CPRA). GDPR indicates eight user rights that are also present in UK's Data Protection Act. CCPA has introduced five main privacy rights, and CPRA two more. The right to opt-out of CCPA has a more limited scope than the right to object in GDPR, as it covers only selling and sharing of personal information, while the right to access includes also in a limited way the right to data portability of GDPR. Therefore, GDPR and CCPA with CPRA share many rights: CPRA has in addition the right to limit use of sensitive personal information and CCPA the right to non-discrimination, while GDPR has in addition the right to restriction of processing, the right to avoid automated decision-making, and more general rights to data portability and to object. Moreover, in both GDPR and CCPA, users need to provide their consent on the collection, usage, and processing of their personal data, and they are typically informed about these aspects via the privacy policy of the system. CCPA gives more emphasis on the possibility to opt-out at a later stage, but removing consent is also feasible in GDPR. We also put an emphasis on cookies in our study, as cookie banners are present in almost all systems and their respect of user preferences is important, as users need to provide their consent for their use~\citep{matte2020cookie}.

\begin{table}[h]
\centering
\begin{adjustbox}{width=0.99\textwidth}
  \begin{tabular}{p{6.9cm}p{6.5cm}}
    \toprule
    \textbf{GDPR (also UK DPA) rights} & \textbf{CCPA/CPRA rights} \\
    \midrule
    Right to information (Articles 13, 14) & Right to know [CCPA]\\
    \midrule
    Right to access (Article 15) & Right to access [CCPA]\\
    \midrule
    Right to rectification (Article 16) & Right to correct (inaccurate personal information) [CPRA]\\
    \midrule
    Right to erasure (Article 17) & Right to delete [CCPA]\\
    \midrule
    Right to restriction of processing (Article 18) & --\\
    \midrule
    Right to data portability (Article 20) & Right to access [CCPA] (\emph{partly covered})\\
    \midrule
    Right to object (Article 21) & Right to opt-out (of sale/sharing of personal information) [CCPA/CPRA] (\emph{partly covered})\\
    \midrule
    Right to avoid automated decision-making (Article 22) & --\\
    \midrule
    \emph{Consent removal equivalent}& Right to opt-out (of sale/sharing of personal information) [CCPA/CPRA]\\
    \midrule
    --& Right to non-discrimination (or Right to equal services and prices) [CCPA]\\
    \midrule
    --& Right to limit use of sensitive personal information [CPRA] \\
    \bottomrule
  \end{tabular}
\end{adjustbox}
\caption{User rights and equivalences in main data privacy laws.}
\label{tab:law-user-rights} 
\end{table}

In addition, a number of principles apply in GDPR with seven principles in Article 5 (listed in Table~\ref{tab:gdpr-principles}). In CCPA, the notion of principles does not formally exist but GDPR principles also apply, e.g., accountability, control, and transparency are central in CCPA, while data minimization and purpose limitation are implicit business obligations. We are thus, using the GDPR terminology for principles in this work.

\begin{table}[h]
\centering
\begin{adjustbox}{width=0.99\textwidth}
  \begin{tabular}{p{3.8cm}p{8.9cm}}
    \toprule
    \textbf{GDPR principle} & \textbf{Short description} \\
    \midrule
    Lawfulness, fairness and transparency & Data processing must be lawful and fair, while transparency is linked with the right to information.\\
    \midrule
    Purpose limitation& Purpose for the data collection needs to be defined clearly.\\
    \midrule
    Data minimization& Minimal data need to be collected and be relevant to the processing purpose.\\
    \midrule
    Accuracy&Personal data need to be accurate and be kept up-to-date.\\
    \midrule
    Storage limitation& Data should not be kept for longer than necessary for the purpose for which they are processed.\\
    \midrule
    Integrity and confidentiality& Includes  relevant security aspects.\\
    \midrule
    Accountability& Data controllers must show compliance with the other legislation principles.\\
    \bottomrule
  \end{tabular}
\end{adjustbox}
\caption{GDPR principles descriptions.}
\label{tab:gdpr-principles} 
\end{table}

\subsection{Rights and principles keywords creation}

In order to be able to capture the above, we created a list of keywords that capture different terminology of the user rights and principles and used it in a subsequent keyword-based search within the issues text to answer RQ1. For the keywords creation, we began with a published list from our prior work containing variations of GDPR, CCPA and CPRA user rights~\citep{kapitsaki2025evolution,kapitsaki2025github_privacy_commits}. The list had been created relying on prior experience of the authors on privacy policies analysis~\citep{kapitsaki2023privacy,kounoudes2024right} and on related work in that area~\citep{tesfay2018privacyguide}. In the current work, we expanded the list following text mining practices~\citep{krippendorff2018content}, considering the different terminology that we found in issues during the initial stages of the current work and all manual investigation of issues that we performed (e.g., \emph{Manual verification \& Precision calculation} step in~\figurename~\ref{fig:methodological-process}). The process was performed in iterations and was continuously refined with new keywords added during the whole process of dataset filtering and manual verification every time a different mention of a user right was encountered. This process of adding new keywords was performed by the two authors that undertook the data curation part of the methodology until the final version of the keywords was produced (both experienced researchers in empirical Software Engineering).

We then expanded the created list using suggestions from OpenAI's ChatGPT (using GPT-5). Using zero-shot prompting, we gave as prompt the list that we created for each user right and asked for additional terms that are developer-oriented or GitHub-specific. We observed that some recommended keywords were too narrow and implementation-specific (e.g., \emph{include metadata in export} provided among the keywords for the right to data portability, \emph{block third-party data sharing} provided among the keywords for the right to restriction of processing) or would already include terms from other keywords (e.g., \emph{non-discrimination} indicated in a longer keyword). One of the authors (the first coder of the manual analysis of RQ1 and RQ2) revised the suggested keywords to remove such cases.

A dedicated list for the principles was also created, using the same approach but with the list created from scratch. The list is more limited, as there is less space for variation for the terminology of some of the principles (lawfulness, fairness and transparency, accuracy, integrity and confidentiality, accountability). For this reason, ChatGPT was used for enrichment only for the following principles: purpose limitation, data minimization, storage limitation. A subset of the terms used for the user rights and the list of all principles (without the terms suggested by ChatGPT) are listed in Table~\ref{tab:rights-principles-examples}. The complete lists and the prompts employed are available in the replication package of the work with 564 and 79 keywords for user rights and principles respectively, while we also created a short list with options for user consent with 12 keywords~\citep{replication}.

\begin{table}[h]
\centering
\begin{adjustbox}{width=0.99\textwidth}
  \begin{tabular}{lp{9.6cm}}
    \toprule
    \textbf{Main term} & \textbf{Additional terms} \\
    \midrule
    Right to opt-out&withdraw consent, remove consent, opt(-)out, do not sell user(s) personal information, do not sell (my/his/her/their) personal information, take back consent, drop consent\\
    \midrule
    Right to erasure & allow deletion, right to delete, right to/of deletion, right of erasure, right to request deletion, right to be forgotten, right-to-be-forgotten, erase information, request erasure, request to erase, erase the personal data, erase (his/her/their/any/user/users) personal data, erase (his/her/their/user/users) data, right to erase, erase (his/her/their/user/users) information, user(s) delete, data delete, deletion right, deleting right, delete right, delete request, delete flow, forget data, forget visitor data\\
    \midrule
    Right to access& right of/to access, right of/to disclosure, right to disclose, access (his/her/their) personal data, access (his/her/their) data, access (his/her/their) personal information, access information, obtain a copy, request a copy, request and receive information, request access, access data, right access, access right, data access, receive data, request (his/her/their) data, request information\\
    \midrule
    \emph{Principles}&lawfulness, lawful, fairness, transparency, transparent, purpose limitation, data minimiz(/s)ation, minimiz(/s)ation of data, collect (the) minimum amount of data, collect minimum data, collection of minimum (amount of) data, accuracy, kept up-to-date, storage limitation, limitation in storage, stored (for) limited time, lawfulness of processing, integrity, confidential(ity), accountability, accountable, control\\ 
    \bottomrule
  \end{tabular}
\end{adjustbox}
\caption{Law user rights subset and principles keywords list.}
\label{tab:rights-principles-examples} 
\end{table}

\subsection{Data collection}

Data collection took place in late 2024 and included all relevant issues until June 2024. Data were collected using the GitHub API, with a search within issues discussions. We focus our analysis on issues, as they are useful items in a repository to plan, discuss, and track work, and have also been used in prior work~\citep{khalajzadeh2022diverse,xiao2022recommending}. We collected only publicly available issues, following the Terms of Service of GitHub and the relevant API terms (Section H)~\citep{githubterms}. The initial size of the dataset is shown in the first column of Table~\ref{tab:dataset-size}.

\subsection{Initial data filtering}

We reduced the size of the dataset via a number of filtering steps explained in this and the next subsections (steps in top part of~\figurename~\ref{fig:methodological-process}).

\textbf{Date filtering.} We removed issues that were created before the respective law's signing date: April 14th, 2016 for GDPR, June 28th, 2018 for CCPA, November 3rd, 2020 for CPRA, May 23rd, 2018 for the UK Data Protection Act, as we are interested only in references to the privacy laws and not other uses of the same abbreviation (\emph{date filter} column in Table~\ref{tab:dataset-size}).

\textbf{Status and user type filtering.} We only considered closed issues, as we are interested in completed discussions (\emph{status filter} column in Table~\ref{tab:dataset-size}). We excluded content created by bots (indicated with \emph{Bot} as user type or having \emph{bot} as part of the username that created the issue), since it cannot give many insights into concerns developers actually have (\emph{user type filter} column in Table~\ref{tab:dataset-size})~\citep{golzadeh2021ground,abdellatif2022bothunter}.

\textbf{Keywords merging.} We removed issues that appeared in the dataset more than once for different keywords of the same law using the issue ID (\emph{merged per law} column in Table~\ref{tab:dataset-size}). We did not use additional filters, e.g., for the language being in English, as even issues in other languages referring to privacy legislation are useful (for the manual issues analysis, we translated such cases to English).

\begin{table}
\centering
\begin{adjustbox}{width=0.99\textwidth}
  \begin{tabular}{p{3.5cm}rrrrrrr}
    \toprule
    \textbf{Keyword} & \multicolumn{6}{c}{\textbf{\# issues}}\\
    \midrule
    & \textbf{collected} & \textbf{date} & \textbf{status}  & \textbf{user} & \textbf{merged} & \textbf{manual \&} & \textbf{final}\\
    &&\textbf{filter}&\textbf{filter} &\textbf{type filter} &\textbf{per law} & \textbf{keywords}\\
    \midrule
    GDPR & 102,192 & 102,042 & 53,671 &36,923& 37,091 & 34,467 & 31,589\\  
    General Data Protection Regulation & 1,558 & 1,551 & 1,018 & 975\\
    \midrule
    CCPA & 5,743 & 5,713& 2,826 & 2,669& 2,689 & 2,306 & 2,063\\
    California Consumer Privacy Act & 264& 261 &134&135&\\  
    \midrule
    CPRA & 234 & 172 & 112 & 109 &112& 59 & 59\\
    California Privacy Rights Act & 55 & 49 & 11 &8\\
    \midrule
    Data Protection Act & 468 & 374 & 227 & 208&208& 185 & 170\\
    \midrule
    \textbf{TOTAL \# repos.} &\textbf{60,005}&\textbf{59,927}&\textbf{23,734}& \textbf{14,123}& \textbf{14,123}  & \textbf{14,055}& \textbf{13,227}\\
    \midrule
    \textbf{TOTAL \# issues} &&&&& \textbf{32,820} &\multicolumn{2}{@{}r}{(\textbf{33,888} with duplicates)}\\
    &&&&&\multicolumn{3}{@{}r}{\textbf{23,063} with law names in title/body}\\
    \bottomrule
\end{tabular}%
\end{adjustbox}
  \caption{Dataset size per keyword.}
  \label{tab:dataset-size}
\end{table}

\subsection{Manual verification and precision calculation}

After using the filters described previously, we performed a manual verification process on a subset of the dataset using the issues in the \emph{merged per law} column in Table~\ref{tab:dataset-size}. This process allowed us to ensure that our dataset contains relevant cases and few false positives. As false positives, we consider issues that make reference to a law name, but are not actually law-relevant discussions. For this purpose, we randomly chose a statistically representative sample of cases from each law using Cochran’s formula~\citep{woolson1986sample}: 742 issues for GDPR, 534 issues for CCPA, and 159 issues for the Data Protection Act. For CCPA and the Data Protection Act, these numbers provide 99\% confidence level for the sample selection with a margin of error ±5\% (assuming a population proportion of 50\%), while for GDPR, 2\% of cases were examined due to the larger number of cases in the dataset (accounting for a confidence level of 99\% with a margin of error ±4.68\% for the same population proportion)~\citep{hazra2017using}. In this process, one of the authors (the first coder of the manual analysis of RQ1 and RQ2) examined the set of issues, but this was performed without examining relevant commits and source code updates, as when analyzing issues for the purpose of answering RQ2. CPRA issues were considered a special case, as the abbreviation CPRA also has other uses (e.g., in the dataset, it is also used for Calculated Panel Reactive Antibody as in~\citep{txmatching968}), so all respective issues were manually examined.  

Based on the manual verification of the samples, we calculated the precision values per law, as listed in Table~\ref{tab:data-manual-precision}, using the following formulation: $precision=\frac{relevant\_retrieved\_cases}{all\_retrieved\_cases}$~\citep{buckland1994relationship}, where we consider relevant issues that refer to privacy laws. In the table, we are also reporting the confidence interval for the proportion of relevant cases. Irrelevant issues for the case of GDPR were indicating other systems' descriptions or texts that included the term GDPR~\citep{ElasticPress3316,matomo-for-wordpress826}. 
For CCPA, among the false positives we found irrelevant cases, where the CCPA law was indicated as part of a job advert~\citep{software-engineering6892}. Cases like this one are difficult to detect without manual analysis. For the Data Protection Act, there were some cases that were referring to laws of other countries (besides the UK DPA). 17 issues had references to Kenya/Rwanda/Uganda (4 cases), India (3 cases), Mauritius, Ghana, Singapore, Canada, Switzerland, Germany, Australia, Denmark, France, PDPA without being more specific (1 case each), while 19 issues were unclear (i.e., it was not feasible to exclude UK DPA or find indication of a specific country). We kept the cases that were not EU-based as relevant, because they still concern privacy law compliance discussions, e.g., in the issue with title: ``\emph{Create legal resource for India (IN) and the Digital Personal Data Protection Act (2023)}"~\citep{openNDS62}. The Data Protection Act keyword is less specific than the other law terms used in the study. For CPRA, a large number of irrelevant cases were found compared to the other laws due to the other uses of the abbreviation. 

\begin{table}
\centering
\begin{adjustbox}{width=0.99\textwidth}
  \begin{tabular}{p{2.4cm}ccccc}
    \toprule
    \textbf{Law} & \textbf{GDPR} & \textbf{CCPA} & \textbf{CPRA} & \textbf{Data Protection Act}& \textbf{TOTAL}\\
    \midrule
    \textbf{Precision}&0.88&0.86&0.53&0.86&0.84\\
    \textbf{99\% Confidence interval}& [0.83, 0.92] &[0.81, 0.91] & [0.53, 0.53] & [0.82, 0.92] & [0.79, 0.89]\\
    \bottomrule
\end{tabular}
\end{adjustbox}
  \caption{Dataset relevance per law.}
  \label{tab:data-manual-precision}
\end{table}

\subsection{Repositories filtering}

Through manual verification, some cases of repositories with irrelevant issues were found, and we decided to remove the respective repository issues from the dataset with the resulting issues in \emph{manual \& keywords} column in Table~\ref{tab:dataset-size}: \emph{rsyslog/rsyslog}, \emph{wazuh/wazuh}, \emph{wazuh/wazuh-qa}, \emph{Cloud-Officer} repositories, \emph{AdguardTeam/AdguardFilters}, \emph{webcompat/web-bugs}, a large number of issues in the repositories of \emph{woocommerce}, \emph{remote-job-boards/software-engineering}, \emph{qiaoyuet/ arxiv\_daily}.
 Details of the reasons for removing these cases are available in the replication package~\citep{replication}.
 
\subsection{BERT fine-tuning and classification, and duplicates removal}

Since the total precision from the manual verification was 0.84, in order to further improve the accuracy of our dataset, we used the manually labeled dataset of the manual verification (as privacy-law relevant or not) to fine tune BERT using \texttt{bert-base-uncased} model (\emph{BERT fine-tuning \& classification} step in~\figurename~\ref{fig:methodological-process})~\citep{devlin2019bert}. We used the maximum length of BERT for tokens (512 tokens) and fine tuned the model using the issues title and body text. Before setting for BERT, we experimented with Llama-3 (1-8B) but the accuracy was very low for this binary classification task (Llama is considered less suitable for classification tasks)~\cite{zhang2025bert}. The accuracy of BERT was very good: $accuracy$=$0.9336$, $precision$=$0.9590$, $recall$=$0.9627$, $f1-score$=$0.9609$. We run the fine-tuned classifier on the filtered dataset that was the result of the manual verification and repository filtering process. The last column in Table~\ref{tab:dataset-size} reflects the final dataset size after the BERT binary classification. Although the final number of issues for CPRA and Data Protection Act is very small, we kept it in our dataset for a more holistic view on privacy legislation. We finally performed \emph{Duplicates removal} for issues that contain indications to more than one law.

\subsection{Data analysis}

\textbf{Overview}. For our qualitative and quantitative analysis, we used statistical analysis with descriptive statistics and thematic analysis. For the dataset overview, we computed descriptive statistics and compared law-relevant and non-law-relevant issues using the Mann-Whitney U test. We performed automated analysis and manual examination on a sample of the final issues dataset (RQ1) and, when necessary, on the commits and changed source code (RQ2). Manual examination allows for a nuanced understanding of the context and subtleties of each issue, while automated analysis enables us to process a large volume of data efficiently. Combining these methods improves the reliability and depth of our findings. Manual analysis was combined with automated analysis also in prior works:~\citep{tufano2024unveiling} manually analyzed 467 commits/pull-requests/issues out of the 1,501 collected, while~\citep{tahaei2020understanding} analyzed a sample of 315 privacy-related questions of Stack Overflow from the 1,733, where topic modeling was applied.

For the manual qualitative data analysis in RQ1 and RQ2, we adopted respectively deductive and inductive thematic analysis~\cite{chandra2019inductive}. In RQ1, the user rights and principles are known, while in RQ2 a bottom-up approach was required to define the categories similar to previous studies~\citep{galster2022soft,tufano2024unveiling,kapitsaki2024exploratory}. Two coders (i.e., two of the authors), both computer scientists experienced in empirical software engineering, participated in the analysis, both with expertise in GDPR and Privacy by Design. One of the coders is an experienced researcher (more than 15 years postdoc experience), whereas the second is a junior researcher (software engineer and PhD candidate). Also three additional experts performed a validation of the created taxonomy of RQ2. Issues in languages other than English were translated using Google Translate with automatic language detection.

\textbf{RQ1 analysis}. \emph{Automated analysis}. For investigating the presence of privacy legislation user rights and law principles in developers' discussions, we performed a keyword search within the issue titles and bodies to see if there are references to specific user rights and principles found in the legislation, using the terms created in the \emph{Rights/principles keywords creation} step of the methodology (subset provided in Table~\ref{tab:rights-principles-examples}). Due to the specificity of the legal terminology, keyword-based search using the created list is more appropriate to answer this RQ compared to other NLP techniques (e.g., topic modeling). We nevertheless, experimented with BERT as described later. We performed the search using all 32,820 issues in the dataset, regardless of where the law name is indicated (in title/body or comments of the issue), as there are many relevant cases where the law name is indicated only in the comments, e.g., in linking repository features with law compliance or referring to cookie banners~\citep{vertx-auth102,slackdump126}. The search was performed in the title/body of each issue.

\emph{Sample selection for manual analysis}. For sample selection for manual analysis, we chose a representative sample (using the \texttt{sample} function in the R programming language). In this process, we used issues that make direct mention to one (or more) of the laws in the title or body of the issue. This ensures that our sample is highly relevant and reduces the likelihood of including unrelated issues. We 
used a sample of 1,186 issues, comprising approximately 5\% of the total of 23,063 issues.

\emph{Issues analysis}, and \emph{Agreement/discussion}. The coders analyzed all issues independently (using a spreadsheet with the available issues), indicated whether the issue included any of the user rights or principles of the laws, and marked the main right or principle when present. A codebook was created for this process that contains explanations for user rights and principles and is available in the replication package of the work~\cite{replication}. Coders spent a total of 2,016 minutes coding (slightly less than 5 working days), with an average of 1.7 minutes per issue. There were 112 cases of disagreement that were resolved via the help of a third author (senior researcher, with four years of postdoc experience) who provided the final category for these cases after inspecting the respective issue. Inter-coder agreement between the two coders was measured using Cohen's kappa ($\kappa=0.846$), indicating substantial agreement~\citep{landis1977measurement}.

We then used this manually labeled dataset to fine-tune BERT, in order to run the classification process on the whole issues dataset. As in the filtering process, we used the maximum tokens of BERT (512 tokens). However, the accuracy was very low ($accuracy$=$0.6653$, $precision$=$0.5067$, $recall$=$0.6653$, $f1-score$=$0.5614$), so we refrain from presenting the results, finding keyword-based search more suitable due to the law-specific terminology that is mixed with other text within the issues.

\textbf{RQ2 analysis}. \emph{Issues/code analysis}, and  \emph{Categorization and taxonomy creation}. Both the issue discussions and relevant commits and source code changes (when applicable) were analyzed. The two coders (same as in RQ1) become familiar with the issues content during the keywords creation, the manual verification of the dataset creation process and RQ1 analysis. The coding categories were developed through a bottom-up iterative process conducted in two rounds. This is in contrast to~\cite{sangaroonsilp2023taxonomy}, where issues from Chrome and Moodle were manually classified among existing categories concerning privacy-related requirements. In the first round, 20\% of the data (222 issues) was analyzed by the two coders, who independently proposed categories with a naming of their choice as they were going through the issues. Each issue could be assigned to one or more categories. After this initial round, the two coders met to review all proposed categories, finalize their naming, and agree on a final set to use for categorizing the remaining elements in the dataset. The initial taxonomy consisted of 22 categories, after resolving any differences in terminology in the category names that referred to the same concepts. In the second round, the remaining issues (80\%) were analyzed. During this second round, issues were placed under the agreed categories but additional categories could be introduced by the two coders and were discussed in the consensus meeting after coding. The two rounds of the coding task took approximately six full working days per coder (2,490 minutes), with an average of 2.1 minutes per issue. 

\emph{Agreement/discussion}. The second coding round was followed by a meeting between the two coders to resolve cases of disagreements in the categorization of all issues from both rounds, along with discussing and deciding final names for additional categories found. There were 297 cases of disagreement. A third author (same third involved researcher as in RQ1) participated in this meeting to assist in agreeing on the final category(/ies) for these cases. The third author analyzed each issue under discussion independently and then discussed with the two coders in order to reach the final category(/ies). The two meetings after each coding round lasted two and six hours respectively. Upon completion of the coding task and before the final consensus meeting, Cohen's kappa was $\kappa=0.717$, indicating substantial agreement.

\emph{Taxonomy validation}. In order to validate the clarity and the completeness of the taxonomy, three external experts among our collaborators were recruited: a technical leader and software engineer, a senior software engineer working on data security and a senior researcher (working on GDPR compliance). A codebook was created after the manual process was completed in order to assist the validation process. The codebook includes definitions of each category, the created clusters and an issue example for each category. The three experts were given access to the codebook and to a short survey. The survey consisted of three parts: 1) provision of comments to the taxonomy to assess the clarity of the categories, 2) possibility to indicate if they believe any relevant category is missing, 3) provision of five issue examples asking them to place them in taxonomy category in order to assess their understanding. The codebook and the short survey given to the experts are available in the replication package~\citep{replication}. For further validation purposes, we collected an unseen dataset of 100 issues and examined manually whether the same categories are applicable.

\section{Results}

\subsection{Dataset overview and contextual comparison}
 
\begin{figure}[!t]
\centering
\includegraphics[scale=0.55]{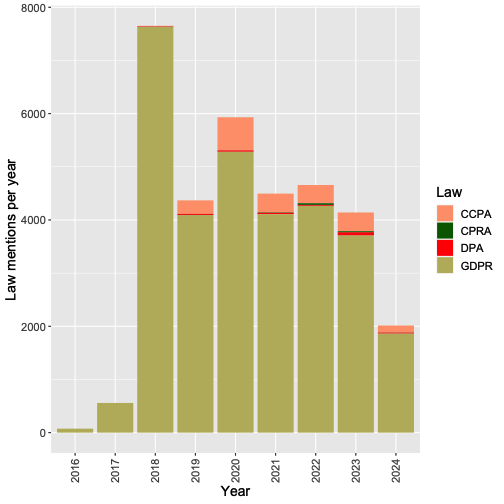}
\caption{Law mentions in issues per year.}
\label{fig:issues-per-year} 
\end{figure}

Issues in the dataset have on average 4.65 comments and remained open for approximately 69 days on average. From the 32,820 issues in the dataset, there are 23,063 issues that mention a data privacy law in their title/body. The yearly distribution of law mentions in the issues of the dataset is depicted in~\figurename~\ref{fig:issues-per-year} with most issues from 2018, where GDPR came into effect, followed by 2020, which is the effect year of CCPA. Some issues make reference to more than one law as would be expected, since repositories need to comply not only with one law (after removing duplicates in Table~\ref{tab:dataset-size}). We compared the dataset with a smaller dataset containing non-law relevant issues (we randomly collected 5,000 issues created between April 2016 and June 2024). After filtering out closed issues to retain only user-created issues, we had 4,160 non-law relevant issues. We used a Mann-Whitney U nonparametric test, since our data on comments and the time discussions remain open do not follow a normal distribution. We found statistically significant differences between law-relevant and non-law relevant issues in terms of the number of comments ($Z=-37.850$) and the number of days the issue remained open ($Z=-41.581$), with $p<0.01$ for both dependent variables (descriptive statistics are shown in Table~\ref{tab:mannwhitney-law-nonlaw}). This observation holds when using a non-GDPR pull request dataset from prior work, after removing duplicates and ensuring that no privacy law is mentioned in the pull requests' title or body in the existing dataset: 15,543 non-law relevant pull requests~\citep{franke2024exploratory}. We argue that comparing with pull requests is valid, as they contain similar data, discuss updates, and are often merged with issues. The Mann-Whitney U test showed statistically significant differences for both comments ($Z=-45.304$) and the number of days the issue remained open ($Z=-51.694$), with $p<0.01$ for both variables. We calculated the effect size using rank-biserial correlation: $r=-0.197$, $r=-0.216$ when comparing with issues and $r=-0.206$, $r=-0.235$ when comparing with pull requests for comments and days of open issues respectively, indicating a small to medium effect according to Cohen's benchmarks~\citep{cohen2013statistical}.

\begin{table}
\centering
\begin{adjustbox}{width=0.75\textwidth}
  \begin{tabular}{llrr}
    \toprule
    \textbf{Variable}&\textbf{Issue type} &\textbf{Mean}&\textbf{Std. deviation}\\
\midrule
\# comments&law-relevant&4.65&30.992\\
&non-law relevant&1.37&5.269\\
\midrule
\# days issue open&law-relevant&69.02&189.235\\
&non-law relevant&38.99&171.609\\
\bottomrule
\end{tabular}
\end{adjustbox}
  \caption{Descriptive statistics between issue discussion types.}
  \label{tab:mannwhitney-law-nonlaw}
\end{table}

\subsection{RQ1. Presence of user rights and principles}

Table~\ref{tab:user-rights-principles-in-issues} reports the findings from the automated analysis, using the list of keywords created in the initial steps of the methodology and enriched with ChatGPT (\emph{Rights and principles keywords creation} step in~\figurename~\ref{fig:methodological-process}) but when comparing the results with the original keywords lists (without the ChatGPT enrichment), we did not observe any differences. For this reason, the replication package makes all keywords lists available (original and ChatGPT enriched)~\cite{replication}. In general, references to specific rights and principles in the title/body are scarce with the right to erasure, the right to opt-out, and the right to access being the most popular rights in discussions, along with consent provision and cookies that appear quite frequently. Among the principles, there is mainly indication to lawfulness, fairness and transparency.

\begin{table}
\centering
\begin{adjustbox}{width=0.99\textwidth}
  \begin{tabular}{llrrrr}
    \toprule
    &\textbf{User right/Principle}  & \textbf{\# issues} &\%& \textbf{\# issues}&\%\\
    & & \textbf{automat.} && \textbf{manual}&\\
    \midrule
    User&Right to erasure&1,137 &3.46\% &102 &9.19\%\\
    rights&Right to opt-out (covering of sale or remove consent) &604 &1.84\% &27 &2.43\%\\
    &Right to access&522 &1.59\% &40 &3.60\%\\
    &Right to rectification&113 &0.34\% &4& 0.36\%\\ 
    &Right to data portability&61 &0.19\% &0 &0\%\\
    &Right to restriction of processing&38 &0.12\% &3 &0.27\%\\
    &Right to information&35 &0.11\% &24 &2.16\%\\
    &Right to object&29 &0.09\% &0 &0\%\\
    &Right to non-discrimination&7 &0.02\% &0 &0\%\\
    &Right to avoid automated decision-making&0 &0\% &1 &0.09\%\\
    &Right to limit use of sensitive personal information&0 &0\% &1 &0.09\%\\
    \midrule
    Other &Consent/opt-in&4,425 &13.48\% &184 &15.51\%\\
    rights&Cookies&4,216 &12.85\% &105 &9.46\%\\
    \midrule
    Law &Lawfulness, fairness and transparency&425 &1.29\% &2 &0.18\%\\
    princi-&Integrity and confidentiality&232 &0.71\% &0 &0\%\\  
    ples*&Accuracy&89 &0.27\% &0 &0\%\\ 
    &Accountability&75 &0.23\% &1 &0.09\%\\  
    &Data minimization&30 &0.09\% &7 &0.63\%\\
    &Storage limitation&4 &0.01\% &12 &1.08\%\\
    &Purpose limitation&3 &0.01\% &1 &0.09\%\\
\bottomrule
\multicolumn{3}{l}{\footnotesize{*Control keyword not examined as being very generic}}\\
\end{tabular}
\end{adjustbox}
  \caption{Presence of legislation user rights and principles in issues.}
  \label{tab:user-rights-principles-in-issues}
\end{table} 

The results of the manual analysis are listed in the last columns of Table~\ref{tab:user-rights-principles-in-issues}. Overall, repositories pay attention to some specific user rights, i.e., right to erasure, right to access, right to opt-out and right to information, but we found scarce references to other rights, such as the right to rectification and only few to law principles. The right to information is very rare in the automated results but is easier to detect via the manual analysis (and hence the percentage is higher), as issues indicate specific updates in privacy policies to include data collection and user rights, without explicitly referring to the right. Handling user's consent for data collection and cookies are the top concern areas found in both automated and manual analysis. During the manual analysis, some false positives were found: 76 issues (6.41\%). The percentages indicated in Table~\ref{tab:user-rights-principles-in-issues} for the manual analysis consider the 1,110 relevant issues.

\noindent\fbox{%
    \parbox{\linewidth}{%
    \textbf{Findings RQ1:} 1) Not many issues make direct reference to legislation user rights and principles. 2) Data collection and cookies consent are the two areas that concern practitioners the most. The right to erasure, the right to access and removing consent (opt-out) are the three most frequent user rights repositories discuss. 3) Principles are scarcely discussed directly, with manual analysis finding storage limitation as the most common.
    }%
}

\subsection{RQ2. Taxonomy of concerns discussed}

\begin{figure}[!t]
\centering
\includegraphics[scale=0.54]{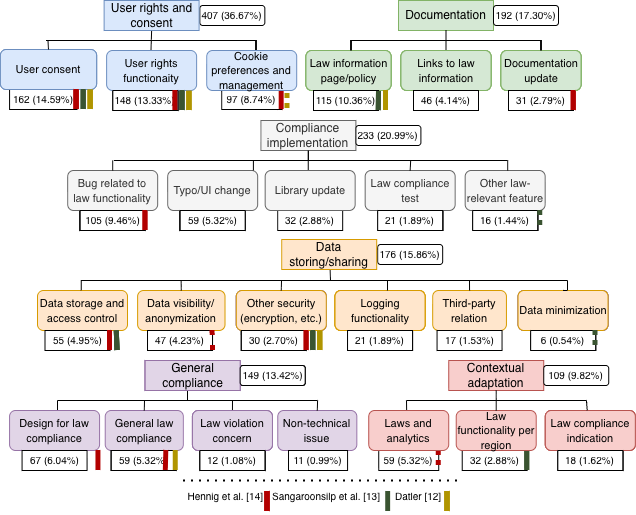}
\caption{Privacy law compliance concerns taxonomy. Color bars on the right of each category refer to overlaps with categories from prior work (dots show partial overlap).}
\label{fig:privacy-concerns-categorization} 
\end{figure} 

The 24 final categories are depicted in~\figurename~\ref{fig:privacy-concerns-categorization}, where they are placed in a taxonomy that was discussed and agreed among all authors. After the initial creation of categories, only two more were added (\emph{Data visibility/anonymization}, and \emph{Library update}) since all new data fit well into the categories already created. The taxonomy consists of six clusters (the two coders created the initial taxonomy and the other two authors made suggestions in terminology before reaching the final version): user rights and consent, compliance implementation, documentation, data storing/sharing, general compliance, and contextual adaptability. The aim of the clustering was to provide more general categories that correspond to the type of activity discussed in the issue and to connect the categories to Privacy by Design. Although most issues were placed under one category, some were placed under two or more (e.g., issue with generic title ``\emph{GDPR Compliance message}" covering both \emph{Links to law information} and \emph{User consent categories}~\cite{icgcargo839}) and for this reason the total occurrences indicated in the figure (in frequencies and percentages underneath each category and above each cluster) are higher than the total number of issues examined (1,110, without the false positives reported in RQ1).

Table~\ref{tab:issues-categories} provides a short description for each category, along with relevant examples (the categories are presented in the table from more popular to less popular). \emph{User rights functionality} and cookies appear fewer times compared to the total occurrence of user rights and cookies, respectively, in RQ1, as some cases are placed in other categories during the detailed categorization (e.g., some cases may indicate user rights but concern primarily bugs or policy updates~\citep{cubecart2338}). In \emph{Other law-relevant feature}, we have added cases not covered in other categories as new features or updates, while general issues of law compliance are addressed in the \emph{General law compliance} category. \emph{User rights and consent} is the most common case when performing updates and is discussed more among practitioners. Issues with data collected for analytics (\emph{Laws and analytics} category) are also related to user consent, while some repositories choose analytics software based on its compliance with privacy laws~\citep{GDPRDialog49}. Other issues discuss \emph{General law compliance} of specific applications, as in~\citep{arx126}. Some issues contain discussions on various topics, as in~\citep{dokuwiki2321}, where contributors discuss all the issues they need to pay attention to due to the introduction of GDPR.

{\footnotesize
  \begin{longtable}{p{2cm}p{5cm}p{4cm}}
    \toprule
    \textbf{Category} & \textbf{Description}& \textbf{Issue title example}\\
    \midrule
User consent&Mechanisms to allow users to provide consent for data handling or relevant preferences, or handle consent&\emph{Show privacy policy and log user consent for GDPR compliance}~\citep{nextcloudregistration1}\\
\midrule
User rights functionality&Functionality related to user rights defined in legislation&\emph{Privacy: Provide a means for an atomic site owner to notify all users of a breach}~\citep{calypso22512}
\\
\midrule
Law information page/policy&Creation/update of privacy policy or law information page, including footers&\emph{Update translations for GDPR Transparency}~\citep{sdk-android153}\\
\midrule
Bug related law functionality&Bugs encountered concerning law compliance related functionality&\emph{Social Share buttons are only available for registered users}~\citep{Kunena-Forum6032}\\
\midrule
Cookie preferences and management&User provides cookies consent/preferences or relevant functionality&\emph{Opt users out of all cookies that are not ``strictly necessary" by default}~\citep{pubpub390}\\
\midrule
Design for law compliance&Considering law compliance when discussing system design/update, including Privacy by Design&\emph{GDPR}~\citep{dokuwiki2321}\\
\midrule
Laws and analytics&Integration and issues with analytics (e.g., Google Analytics, tracking, ads)&\emph{Localytics \& Adobe}~\citep{GDPRDialog49}\\
\midrule
General law compliance&General concerns for law compliance or general compliance functionality&\emph{EU General Data Protection Regulation compliance support?}~\citep{arx126}\\
\midrule
Typo/UI change&Changes in UI, including fixes in wording and typos&\emph{Add GDPR (dark) design dialog before TOS}~\citep{spectre402}\\
\midrule
Data storage and access control&Storage of data, including data retention duration and controlling access&\emph{Old users removed}~\citep{ceda-services-portal2}\\
\midrule
Data visibility/anonymization&Defining which data are visible and which are/become anonymized&\emph{How to anonymize 'Ip Address' dimension}~\citep{Open-Web-Analytics683}\\
\midrule
Links to law information&Adding or correcting links to law pages, e.g., to privacy policy, to GDPR&\emph{``Privacy policy" is not listed in footer}~\citep{couchers2399}\\
\midrule
Law functionality per region&Differentiations in code relevant to law applicability regions, e.g., EU, US&\emph{CCPA For California residents)~\citep{poingstudios81}}\\
\midrule
Library update&Update to a new library version or library change&\emph{Remove gatsby-plugin-gdpr-tracking from OSS Site}~\citep{newrelic734}\\
\midrule
Documentation update&Updating system documentation to reflect  law compliance&\emph{Added GDPR api}~\citep{sdk-sample-code-ios2}\\
    \midrule
Other security (encryption etc.)&Application of any security mechanism&\emph{Encrypt \& Decrypt Emails In The Database}~\citep{django-newsfeed8}\\
    \midrule
Logging functionality&Logging mechanisms in the system including for auditing purposes&\emph{Optional console logging}~\citep{MultiChat70}\\
\midrule
Law compliance indication&Law compliance indication additions/updates in source code&\emph{GDPR Update}~\citep{veronalabs38}\\
\midrule
Law compliance test&Testing related to law compliance features&\emph{GDPR Compliance Assessment UX Review}~\citep{uradotdesign24}\\
\midrule
Third-party relation&Relation with third party providers or software&\emph{Provide an ability to login with third party services}~\citep{re-cite35}\\
\midrule
Other law-relevant feature&Adding new functionality for law compliance not covered in other categories&\emph{Expose `get-gdpr-utils` as appropriate}~\citep{gebruederheitz8}\\
\midrule
Law violation concern&Raising concerns for law violations (existing or potential)&\emph{Checkout is loading Google Fonts (Potential for GDPR Violation?)}~\citep{Adyen2403}\\
\midrule
Non-technical issue&Any non-technical issue discussed&\emph{Authorisation: readonlyaccess to filled in DMP/GDPR report by other groups~\citep{DMPonline_v427}}\\
\midrule
Data minimization&Ensuring that only required data are collected, as in law principles&\emph{Only require e-mail for b2c booking}~\citep{open-booking-api99}\\
    \midrule
Many categories& User rights functionality, Laws and analytics & \emph{Google Analytics GDPR}~\citep{city-grind13} \\
& User rights functionality, Cookie preferences and management & \emph{revoking consent?}~\citep{AspNetCoreDocs8174}\\
& Logging functionality, Data visibility/anonymization & \emph{Option for masking user name.}~\citep{IdentityServer43841}\\
\bottomrule
    \caption{Description of privacy law-related concerns discussed in GitHub issues.} 
  \label{tab:issues-categories}
\end{longtable}
}

Other cases discuss logging that includes user activity data, as in~\citep{MultiChat70}. As many GitHub issues refer to bugs, we have also encountered many discussions and fixes on \emph{Bugs related with law functionality}. Relevant documentation is also a concern to reflect privacy compliance activities (\emph{Documentation update} category), and especially creation and updates to privacy policies and other law-relevant pages, or links to them (\emph{Law information page/policy} and \emph{Links to law information} categories), which are also relevant to the right to information. Repositories are also making changes to mark \emph{Law compliance indication} for the system, whereas \emph{Law compliance test}s for relevant law features are also found in discussions and code changes. The enforcement of privacy laws is closely related to the existence of security mechanisms that support privacy, such as encryption, but these appear less frequently among the issues discussed (2.70\% of the issues examined). \emph{Data visibility/anonymization} that is also relevant to security practices is slightly more frequent in the discussions (4.23\%). A very interesting category has been created from the discussions that cover contextual conditions, adapting \emph{Law functionality per region}. Some categories are less frequent and do not concern the majority of repositories (\emph{Third-party relation}, \emph{Law violation concern} and \emph{Data minimization}), even though they all cover important areas. There are also some less potentially important concerns, such as \emph{Typo/UI change}s, \emph{Library update} issues and \emph{Non-technical issue} discussions.

\begin{table}
\centering
\begin{adjustbox}{width=0.88\textwidth}
  \begin{tabular}{llrr}
    \toprule
    \textbf{Variable}&\textbf{Issue category} &\textbf{Mean} &\textbf{SD}\\
    \midrule
\# comments&User consent&2.40&9.821\\
&User rights functionality&2.00&3.753\\
&Law information page/policy&1.55&2.735\\
&Bug related to law functionality&3.29&.5.092\\
&Cookie preferences and management&1.68&3.338\\
&Design for law compliance&\textbf{3.70}&5.518\\
&Laws and analytics&2.21&2.858\\
&General law compliance&1.02&2.712\\
&Typo/UI change&1.71&2.508\\
&Data storage and access control&1.47&2.676\\
&Data visibility/anonymization&2.97&4.433\\
&Links to law information&0.85&1.986\\
&Law functionality per region&1.31&1.401\\
&Library update&1.04&1.846\\
&Documentation update&1.04&1.567\\
&Other security (encryption, etc.)&2.27&4.229\\
&Logging functionality&1.89&3.296\\
&Law compliance indication&0.47&0.834\\
&Law compliance test&3.05&4.968\\
&Third-party relation&3.00&2.944\\
&Other law-relevant feature&1.31&2.243\\
&Law violation concern&\textbf{4.29}&4.645\\
&Non-technical issue&1.36&2.063\\
&Data minimization&1.33&0.577\\
\midrule  
\# days issue open&User consent&59.33&222.475\\
&User rights functionality&86.32&257.517\\
&Law information page/policy&24.20&105.455\\
&Bug related to law functionality&75.82&200.936\\
&Cookie preferences and management&52.70&162.560\\
&Design for law compliance&\textbf{193.93}&353.348\\
&Laws and analytics&37.41&80.907\\
&General law compliance&25.17&76.091\\
&Typo/UI change&49.63&174.354\\
&Data storage and access control&67.49&208.528\\
&Data visibility/anonymization&68.26&189.411\\
&Links to law information&37.36&132.851\\
&Law functionality per region&56.81&163.000\\
&Library update&9.04&17.012\\
&Documentation update&41.64&105.910\\
&Other security (encryption, etc.)&21.15&54.297\\
&Logging functionality&11.33&25.962\\
&Law compliance indication&45.33&143.365\\
&Law compliance test&13.95&23.648\\
&Third-party relation&\textbf{178.14}&282.278\\
&Other law-relevant feature&12.63&38.110\\
&Law violation concern&34.86&31.924\\
&Non-technical issue&18.27&29.080\\
&Data minimization&21.67&34.933\\
\bottomrule
\end{tabular}
\end{adjustbox}
\caption{Descriptive statistics for comparison of issue categories.}
  \label{tab:categories-comparison}
\end{table}

\textbf{Taxonomy validation results}. The experts completed the survey and discussion with each of them followed afterwards. They confirmed that the categories are generally clear and well-structured. One expert suggested adding Privacy by Design (or Architectural decisions) as an additional category, but during the subsequent discussion with the expert, it was decided that this is already covered in a category we had already devised (\emph{Design for law compliance}) and in the categories grouping clusters. Privacy by Design was thus added in the definition of the category. Another expert suggested some adaptations in the descriptions that we adopted in the presented version (e.g., \emph{cookie preferences} instead of \emph{cookies preferences}, relating \emph{logging functionality} with audits). The same expert suggested that anonymization falls under security practices (for this reason we added the word \emph{other} in the security category as it was initially missing), and that data access should also be present (for this reason we added access control together with data storage, as in the initial version only data storage was indicated). The experts placed almost all of the provided issues under the categories that were assigned in our coding process apart: one expert placed~\citep{aptible16} under \emph{Design for law compliance}, whereas it refers to \emph{Law information page/policy} in our coding (as the file changes in the respective commit indicates). We regarded the general result as indicator that the experts gained a good understanding of the taxonomy.

Moreover, we randomly used 100 closed and user-created collected issues from April 2026 that were classified as law-relevant via the fine-tuned BERT model. This sample size was chosen to achieve thematic saturation, a point where subsequent issues analysis does not provide additional categories (achieved after analyzing 76 issues). Prior empirical studies in software engineering similarly used samples of this scale~\cite{miller2022did}. The analysis task was completed by one of the authors. We found that all analyzed issues fit into the categories of the taxonomy with the following top 5 categories encountered: \emph{User consent} (21 cases), \emph{User rights functionality} (17 cases), \emph{Logging functionality} (11 cases), \emph{Data visibility/anonymization} (10 cases), \emph{Design for law compliance} (8 cases) (frequencies available in the replication package~\citep{replication}). Among those, we found five cases that are relevant to creating skills for privacy law compliance for generative AI coding tools or to prompts and these were placed under the \emph{Design for law compliance} category~\cite{trello-api-skill,moral-core22}.

\begin{figure}[!t]
\centering
\includegraphics[scale=0.57]{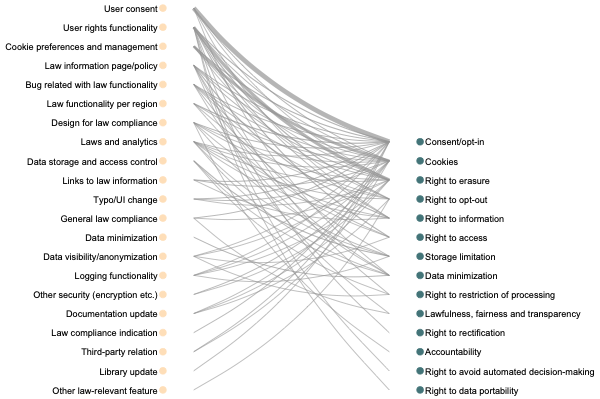}
\caption{Relation between taxonomy categories and privacy law rights/principles.}
\label{fig:categories-to-rights-principles} 
\end{figure}

\textbf{Categories and user rights/principles.}~\figurename~\ref{fig:categories-to-rights-principles} illustrates the relation between the taxonomy categories identified in RQ2 and the privacy-law user rights and principles examined in RQ1. This helps clarify how taxonomy categories operationalize specific legal requirements, showing that the taxonomy is connected to user rights and principles even though it was derived bottom-up from developers’ discussions. The strongest links are concentrated around a limited number of categories and user rights, especially between \emph{User consent} category and Consent/opt-in user right, \emph{Cookie preferences and management} category and Cookies, and \emph{User rights functionality} category and rights, such as right to erasure, right to opt-out, and right to access.

\textbf{Categories comparison.} We compared the categories to determine if there are differences in the number of comments or the time an issue remains open, depending on its category. To analyze this, we conducted a Kruskal-Wallis H rank-based non-parametric test, using the number of comments or the days an issue remained open as the dependent variables and the issue category (from the manual analysis) as the independent variable. For this test, we used the 969 issues that belong to exactly one category, as each case must belong to only one group. The results were statistically significant for both dependent variables ($\chi^2(2)=82.588$, $p<0.01$ for comments; $\chi^2(2)=77.618$, $p<0.01$ for the days the issue remained open), with descriptive statistics shown in Table~\ref{tab:categories-comparison}. The effect size, which we calculated using eta squared, was $\eta^2=0.055$ for the number of comments and $\eta^2=0.050$  for the number of days, indicating effect close to medium according to Cohen’s benchmarks~\citep{cohen2013statistical}.

Discussions on law violations and design for law compliance tend to generate more comments, indicating the importance of clearly defining necessary actions when designing privacy features. Design discussions remain open longer than other categories, possibly due to the need for more reflection or because they might be further discussed in other issues specific for the implementation of the respective functionality and be closed after the implementation is completed, whereas potential violations do not stay open long. This pattern also applies to security-related discussions and library updates, among others. Third-party relations tend to stay open longer than other categories, requiring more time to resolve due to involving external parties, and they also receive relatively more comments.

\noindent\fbox{%
    \parbox{\linewidth}{%
    \textbf{Findings RQ2:} 1) Practitioners discuss features/bugs, consent related issues, documentation, data storing/sharing, adaptability and general compliance, when it comes to privacy law compliance. 2) User consent related and user rights are the types of functionalities usually discussed and added in GitHub repositories, followed by indications of bugs related with law-relevant activities and cookie management and relevant user preferences. 3) Law violation concerns and design for law compliance trigger more comments, while issues for design for law compliance and third-party relations remain open longer.
    }%
}

\section{Discussion}

\subsection{Findings}

Our results indicate that practitioners are attentive to specific aspects, with user consent provision being the most frequently discussed topic. It is the top category in the categorization performed in RQ2 and is the right with the highest presence in both automated and manual analysis of RQ1. Users must give their consent for data collection and the privacy policy when registering or using a system for the first time. Cookie consent, requested typically via cookie banners, is also popular and captured in another category in our taxonomy accounting for 8.74\% of the issues examined manually -- as a more specific case of data collection -- but is highly related with general user consent actions. Documentation-relevant updates that include primarily the privacy policy is the top third cluster in our categorization (RQ2). Some important issues, such as security mechanisms other than anonymization present in 30 issues of our manual analysis (RQ2), are not discussed as extensively as we might have expected. This suggests that these concerns are addressed separately from law compliance, and thus were not captured in our dataset as they contain security-relevant terminology (which we did not use for data collection purposes). Still, these issues tend to remain open for fewer days compared to other categories because of their urgency (21.15 days with five other categories out of the 24 remaining open for an even shorter period). The manual validation of the taxonomy showed that there are cases of providing instructions to AI agents, indicating that using AI coding tools for compliance purposes is an emerging area.

Direct reference to legislation principles in issues was scarce both in the automated and manual analysis of RQ1, which may be attributed to the fact that difference terminology may be used when they are discussed (e.g., we found cases of storage limitation in the \emph{Data storage and access control)} category of our taxonomy~\citep{ceda-services-portal2}). The connection between categories of RQ2 and user rights/principles of RQ1 indicates that developers most often discuss privacy-law compliance through implementation-oriented concerns rather than through explicit legal terminology.

As some prior work has examined privacy-relevant requirements, we compared our work with those works in Table~\ref{tab:groups-prior-work}. The main categories in the issues created from GitHub issues in~\cite{hennig2026whos} evolve around feature requests, existing features and bugs that are related with privacy, security, consent and regulation compliance, that are more generic compared to our more detailed categorization. The two categorizations are not contradictory nevertheless. Concerning the categorization of privacy-relevant developers requirements that was created by analyzing privacy laws and frameworks (GDPR, PDPA of Thailand, APEC) by~\cite{sangaroonsilp2023taxonomy}, the requirements in that work extend beyond software systems and are thus more procedural than in our taxonomy. Comparing our work with that of~\cite{datler2023intended}, who grouped issues into five categories, we find that the main categories (consent and user rights) are present in both studies.  All four studies make reference to security concerns but compared to the existing three works, we are reflecting more specific software development relevant concerns discussed by the OSS contributors (e.g., \emph{Logging functionality}, and \emph{Law compliance test} are examples of such specific categories not found in prior works). For a more detailed comparison between categories, color bars on the right of each category in~\figurename~\ref{fig:privacy-concerns-categorization} show overlaps with categories from prior work (dots show only partial overlap). Nine of our categories have not been captured in prior work, while four are only partially captured.

\begin{table}
\centering
\begin{adjustbox}{width=0.97\textwidth}
  \begin{tabular}{p{3.8cm}p{4.3cm}p{2.6cm}p{2.6cm}}
    \toprule
    \textbf{Clusters/categories in our work} & \textbf{Issues in~\cite{hennig2026whos}} & \textbf{Categories in~\cite{sangaroonsilp2023taxonomy}} & \textbf{Groups in~\cite{datler2023intended}} \\
    \midrule
Features: Bug related law functionality&Bug with implication for privacy, Bug with implication for consent interaction&--&-- \\
\midrule
Features: (mainly) User rights functionality, Other law-relevant feature& Feature request for privacy enhancement/regulation compliance, Existing feature with implication for privacy&User participation, Complaint/request, Breach&Data subject control rights (included) \\
\midrule
Consent related: (mainly) User consent, Cookie preferences and management & Feature request for consent interaction, Cookie banner missing&Notice, User desirability&Notice and consent\\
\midrule
Documentation& Feature request for privacy documentation&Notice&Notice and consent\ \\
\midrule
Data storing/sharing& Request for information about stored data, Private information disclosed in discussion&Data processing&\emph-- \\
\midrule
Adaptability: (mainly) Law functionality per region, Design for law compliance& Feature request with implication for privacy/security& Local requirements (R65) under Security category&-- \\
\midrule
General compliance: Other security (encryption, etc.)& Feature request for security enhancement, Bug with implication for security, Existing feature with implication for security&Security & Data security \\
\midrule
General compliance: other categories & Other privacy-related issues, Compliance evaluation&Data processing& General applicability \\
\midrule
--&--&--&Organizational requirement \\
\bottomrule
\end{tabular}
\end{adjustbox}
  \caption{Comparison of RQ2 results with prior work.}
  \label{tab:groups-prior-work}
\end{table}

\subsection{Implications}

\subsubsection{Implications for practitioners} 

\textbf{Paying attention to all compliance areas}. Practitioners should ensure that all issues commonly discussed (such as user consent, rights, and cookies) are addressed in their systems, but they should also give attention to specific user rights, which are less frequently discussed or sometimes not discussed at all according to the results of RQ1 (e.g., right to restriction of processing, right to object, right to avoid automated decision-making). Although rights, such as the right to erasure and the right to access, are straightforward, there are less clear rights, such as the right to restriction of processing and the right to data portability. More attention also needs to be paid to core security mechanisms, which are a requisite for privacy compliance but are not often discussed, as they appear only in 2.70\% of the issues manually examined in RQ2~\citep{granata2024gdpr}. Nevertheless, security cases might have also been discussed as aforementioned elsewhere or treated offline.

\textbf{Making available guides on compliance}. Currently, there is no central location where developers can find guidance on privacy compliance. Even though each software system has its unique needs, there are common items (e.g., user rights) that organizations such as GitHub could make available as good practices for privacy compliance, with the repository contributors' consent. Some systems make available relevant plugins for compliance (e.g., for WordPress, as we found among the issues examined manually in RQ2~\citep{mailchimp-for-wordpress556}). Prior work has indicated a need for more tools to support compliance, as developers often place significant trust in enforcement by mobile application markets~\citep{alomar2022developers}.

\textbf{Providing solutions for compliance for popular areas of discussion}. We are providing a number of examples of how popular compliance areas are resolved in GitHub issues that we examined in RQ2, while future work is required for an exhaustive list: 
\begin{itemize}
    \item \textbf{User consent:} Among other user consent functionalities, the software system should ensure there is user consent for marketing emails, even for guest visitors (example issue title: ``\emph{CHECKOUT-4640 Allow guest shopper to provide marketing email consent}"~\citep{bigcommerce222}).
    \item \textbf{User rights functionality:} For the right to erasure, there are different options on how to implement it, for instance, administrators can delete users data when requested (example from issue body: ``\emph{As an IT-Administrator I need to delete files according to GDPR. In a lot of cases it is required to make sure that a deleted file cannot be restored, so I need a way to securely erase files in Powershell.}"~\citep{microjam97}), or end-users can have the possibility to delete specific entries they had added in the past (example issue title: ``\emph{[Feature] Delete a post}"~\citep{Human-Connection330}).
    \item \textbf{Law information page/policy:} A privacy policy needs to be in place and among the information it should contain, it needs to have a clear statement about how and why data are processed (example issue title: ``\emph{Write privacy statement}"~\citep{Artaxerxes85}), while it would be convenient if it is available in different languages, so translation support is needed (example issue title: ``\emph{Make all GDPR pages translatable}"~\citep{matomo15574}).
\end{itemize}

\subsubsection{Implications for educators}

\textbf{Creating educational material on areas less commonly discussed.} Our work (to answer RQ1) found that specific user rights are frequently discussed in issues, indicating a need for educational material on those common rights, but also on rights that are less frequent, such as the right to rectification. It would be beneficial to include privacy law compliance in software engineering courses, as similarly suggested for law students concerning technology education~\citep{volini2019perspective}.

\textbf{Policy makers offering guidance on law compliance.} The insight on the categorization of the concerns raised in RQ2 is also relevant for policymakers who can assist in clarifying specific procedures for law compliance. According to our contextual analysis, law-relevant issues received more comments than non-law relevant issues and these issues also tended to remain open longer. This suggests that law-relevant changes are more challenging and may require more time to reach consensus among contributors and to be appropriately handled in source code changes.

\subsubsection{Implications for researchers}
\textbf{Creating automated tools and recommender systems for law compliance in source code.} The categories that emerged from our work in RQ2 highlight the need for automated mechanisms to target those specific categories (e.g., anonymizing data as in \emph{Data visibility/anonymization} category, or for fixing law-relevant types as in \emph{Typo/UI change} category). Recommendations for implementing privacy law compliance have also been emphasized in previous work~\citep{alomar2022developers}. Research aimed at suggesting appropriate methods for handling law-relevant concerns discussed in issues is needed so that relevant tools can be integrated into GitHub and other issue tracking systems.

\textbf{Making law compliance faster.} Accelerating the law compliance process is also an important research direction, as our results show that these tasks take longer to handle (compared to non-law relevant issues and pull requests). AI coding tools can assist in this direction, as they may help developers in exploring implementation alternatives or drafting compliance-related functionality more quickly. They cannot though be relied upon alone to guarantee secure or legally compliant outcomes~\citep{ferreyra2025good}. According to prior work, there is currently a very small number of developers employing generative AI tools in order to meet the privacy requirements of the software they are developing, and a lot of manual effort also needs to be put to ensure these requirements are respected~\citep{madampe2025we}. Policy as Code tools, such as RuleKeeper that assists in automatic enforcement of GDPR, can also be considered in combination with the results of our work~\citep{ferreira2023rulekeeper}.

\section{Threats to validity}

\emph{External validity}. We collected issues from GitHub, but repositories with external issue tracking systems were not included. However, GitHub hosts a vast number of OSS repositories across various domains, so we believe our findings are broadly representative of general developer practices related to privacy law compliance in the open source community. Concerning proprietary software, as we did not study industrial issue tracking systems, different privacy concerns may arise as most popular. We also did not consider other sources where privacy legislation might be addressed, such as pull requests or code comments. This is mitigated by the fact that issues serve as a primary discussion space on GitHub, where key compliance concerns are likely to be raised and documented. Future work could, however, focus on these aspects as well. Our manual analysis was restricted to a sample of the dataset, and the relevant percentages (mainly for the categories that were only manually assigned) may differ when considering the entire dataset or sources outside of GitHub. Nevertheless, the sample size used was statistically representative (5\% of the whole dataset) and thus captures the main trends effectively. 

\emph{Internal Validity}. We do not expect our work to be affected by internal validity issues, as we relied on commonly used APIs and libraries for data collection and analysis (of R and Python programming languages). The GitHub API is the official way for retrieving up-to-date data from repositories. The manual effort put in the filtering steps may hinder reproducing those steps of the process.

\emph{Construct Validity}. Even after careful filtering and extensive manual verification, the dataset contains some false positives, as found in the sample used during the manual analysis for RQ1. We primarily used keyword-based and manual filtering to emphasize human analysis alongside BERT binary classification, while we experimented with BERT for RQ1 and RQ2 and obtained low accuracy as it seems to miss law-specific terminology. In the manual analysis, we examined each issue individually, which may have affected the categorization in cases where an issue is linked to other issues. Despite the filtering measures during the dataset construction, we cannot determine whether the initial data collection omitted relevant cases from GitHub, as this is impossible to control or verify. The generic Data Protection Act keyword with a few occurrences of issues from jurisdictions other than EU and US, gave a slightly broader scope to the study. Given also our extensive verification, missed cases are unlikely to affect the overall results.

\emph{Conclusions Validity}. In some cases, the exact concern of the issue is not clear, as it may not be explicitly stated (e.g., data minimization might be indirectly addressed by collecting minimal user data), which may have affected the categorization in RQ2. In general, the work relied heavily on the manual coding of the authors, which may be prone to human error. We mitigated this threat by performing the manual analysis in iterations (in RQ2), using two main coders along with a third author for cases of disagreement (in RQ1 and RQ2), and combining it with automated analysis. Moreover, the categorization is based on developers' views. A fully user rights/principles-based taxonomy would require a different, more normative analysis. Another potential threat is the size of issues selected for taxonomy verification purposes (100 issues). To assess whether another sample size would affect the findings, we performed a stability check. We observed that no new categories from the taxonomy emerged after analyzing 76 issues. Increasing the sample size to 100 did not affect the findings.

\section{Conclusions}

We analyzed 32,820 GitHub issues from 13,227 repositories related to privacy law compliance, focusing on major privacy laws, such as GDPR, CCPA, CPRA, and the Data Protection Act. Our analysis revealed that the most frequently discussed user rights are the rights to erasure, access, and opt-out. Through manual analysis of a subset of the issues, we identified 24 categories of developer concerns that we grouped into six clusters: user rights and consent, compliance implementation, documentation, data storing/sharing, general compliance, and contextual adaptability. Our future work will further validate our findings through extended surveys and interviews with contributing software engineers to complement the existing taxonomy validation. We also plan to explore the potential of LLMs for handling main privacy development concerns, and will examine suitable mechanisms for achieving data privacy compliance. We will finally investigate Policy as Code mechanisms to integrate solutions for privacy concerns in the framework of Continuous Integration~\citep{ferreira2023rulekeeper}. 

\vspace{5mm} 

\textbf{Acknowledgements} We would like to thank Sotiris Diamantopoulos, Savvas Savvides, and Evangelia Vanezi who performed the validation of the privacy law compliance taxonomy. 

\vspace{5mm} 

\textbf{Funding}. This research did not receive any specific grant from funding agencies in the public, commercial, or not-for-profit sectors.

\vspace{5mm} 

\textbf{Competing interests}. The authors declare that they have no known competing financial interests or personal relationships that could have appeared to influence the work reported in this paper.

\vspace{5mm} 

\textbf{Data statement.} All data collected and used in this study, including the datasets and software, are available at the following repository~\citep{replication}, whereas the fine-tuned BERT model is available on HuggingFace: \url{https://huggingface.co/gkapi/bert-uncased-privacy-law-binary-model}.

\vspace{5mm} 

\textbf{CRediT authorship contribution statement}. \textbf{Georgia M. Kapitsaki:} Conceptualization, Investigation, Methodology, Data curation, Formal analysis, Software, Resources, Writing – original draft, Writing – review and editing. \textbf{Maria Papoutsoglou:} Data curation, Formal analysis, Validation, Writing – review and editing. \textbf{Christoph Treude:} Methodology, Validation, Writing – review and editing. \textbf{Ioanna Theophilou:} Formal analysis, Writing – review and editing.

\bibliography{sample-base}

@String{Computing = "Computing" }

@String{Springer = "Springer-Verlag" }

@inproceedings{khalajzadeh2022diverse,
  title={How are diverse end-user human-centric issues discussed on GitHub?},
  author={Khalajzadeh, Hourieh and Shahin, Mojtaba and Obie, Humphrey O and Grundy, John},
  booktitle={Proceedings of the 2022 ACM/IEEE 44th International Conference on Software Engineering: Software Engineering in Society},
  pages={79--89},
  year={2022}
}

@inproceedings{xiao2022recommending,
  title={Recommending good first issues in GitHub OSS projects},
  author={Xiao, Wenxin and He, Hao and Xu, Weiwei and Tan, Xin and Dong, Jinhao and Zhou, Migahui},
  booktitle={2022 IEEE/ACM 44th International Conference on Software Engineering (ICSE)},
  pages={1830--1842},
  year={2022},
  organization={IEEE}
}

@inproceedings{miller2022did,
  title={“Did You Miss My Comment or What?” Understanding Toxicity in Open Source Discussions},
  author={Miller, Courtney and Cohen, Sophie and Klug, Daniel and Vasilescu, Bogdan and K{\"a}stner, Christian},
  booktitle={In 44th International Conference on Software Engineering (ICSE’22)},
  year={2022}
}

@article{voigt2017eu,
  title={The eu general data protection regulation (gdpr)},
  author={Voigt, Paul and Von dem Bussche, Axel},
  journal={A Practical Guide, 1st Ed., Cham: Springer International Publishing},
  volume={10},
  number={3152676},
  pages={10--5555},
  year={2017},
  publisher={Springer}
}

@inproceedings{vanezi2019gdpr,
  title={GDPR Compliance in the Design of the INFORM e-Learning Platform: a Case Study},
  author={Vanezi, Evangelia and Kouzapas, Dimitrios and Kapitsaki, Georgia M and Costi, Theodora and Yeratziotis, Alexandros and Mettouris, Christos and Philippou, Anna and Papadopoulos, George A},
  booktitle={2019 13th International Conference on Research Challenges in Information Science (RCIS)},
  pages={1--12},
  year={2019},
  organization={IEEE}
}

@article{li2021developers,
  title={How developers talk about personal data and what it means for user privacy: A case study of a developer forum on reddit},
  author={Li, Tianshi and Louie, Elizabeth and Dabbish, Laura and Hong, Jason I},
  journal={Proceedings of the ACM on Human-Computer Interaction},
  volume={4},
  number={CSCW3},
  pages={1--28},
  year={2021},
  publisher={ACM New York, NY, USA}
}

@inproceedings{bissyande2013got,
  title={Got issues? who cares about it? a large scale investigation of issue trackers from github},
  author={Bissyand{\'e}, Tegawend{\'e} F and Lo, David and Jiang, Lingxiao and R{\'e}veillere, Laurent and Klein, Jacques and Le Traon, Yves},
  booktitle={2013 IEEE 24th international symposium on software reliability engineering (ISSRE)},
  pages={188--197},
  year={2013},
  organization={IEEE}
}

@article{alomar2022developers,
  title={Developers Say the Darnedest Things: Privacy Compliance Processes Followed by Developers of Child-Directed Apps},
  author={Alomar, Noura and Egelman, Serge},
  journal={Proceedings on Privacy Enhancing Technologies},
  volume={4},
  number={2022},
  pages={24},
  year={2022}
}

@inproceedings{tahaei2020understanding,
  title={Understanding privacy-related questions on Stack Overflow},
  author={Tahaei, Mohammad and Vaniea, Kami and Saphra, Naomi},
  booktitle={Proceedings of the 2020 CHI conference on human factors in computing systems},
  pages={1--14},
  year={2020}
}

@inproceedings{parsons2023understanding,
  title        = {Understanding Developers Privacy Concerns Through Reddit Thread Analysis},
  author       = {Jonathan Parsons and Michael Schrider and Oyebanjo Ogunlela and Sepideh Ghanavati},
  booktitle    = {Joint Proceedings of REFSQ-2023 Workshops, Doctoral Symposium, Posters \& Tools Track, and Journal Early Feedback Track},
  series       = {CEUR Workshop Proceedings},
  volume       = {3378},
  year         = {2023},
  url          = {https://ceur-ws.org/Vol-3378/NLP4RE-paper5.pdf},
  note         = {© 2023 CC BY 4.0}
}

@article{goldman2020introduction,
  title={An introduction to the california consumer privacy act (ccpa)},
  author={Goldman, Eric},
  journal={Santa Clara Univ. Legal Studies Research Paper},
  year={2020}
}

@inproceedings{diepenbrock2023analysis,
  title={An Analysis of Stack Exchange Questions: Identifying Challenges in Software Design and Development with a Focus on Data Privacy and Data Protection},
  author={Diepenbrock, Andreas and Fleck, Jonas and Sachweh, Sabine},
  booktitle={Proceedings of the 18th International Conference on Availability, Reliability and Security},
  pages={1--7},
  year={2023}
}

@inproceedings{perera2019towards,
  title={Towards integrating human values into software: Mapping principles and rights of GDPR to values},
  author={Perera, Harsha and Hussain, Waqar and Mougouei, Davoud and Shams, Rifat Ara and Nurwidyantoro, Arif and Whittle, Jon},
  booktitle={2019 IEEE 27th international requirements engineering conference (RE)},
  pages={404--409},
  year={2019},
  organization={IEEE}
}

@inproceedings{matte2020cookie,
  title={Do cookie banners respect my choice?: Measuring legal compliance of banners from iab europe’s transparency and consent framework},
  author={Matte, C{\'e}lestin and Bielova, Nataliia and Santos, Cristiana},
  booktitle={2020 IEEE Symposium on Security and Privacy (SP)},
  pages={791--809},
  year={2020},
  organization={IEEE}
}

@inproceedings{galster2022soft,
  title={What soft skills does the software industry* really* want? An exploratory study of software positions in New Zealand},
  author={Galster, Matthias and Mitrovic, Antonija and Malinen, Sanna and Holland, Jay},
  booktitle={Proceedings of the 16th ACM/IEEE International Symposium on Empirical Software Engineering and Measurement},
  pages={272--282},
  year={2022}
}

@article{buckland1994relationship,
  title={The relationship between recall and precision},
  author={Buckland, Michael and Gey, Fredric},
  journal={Journal of the American society for information science},
  volume={45},
  number={1},
  pages={12--19},
  year={1994},
  publisher={Wiley Online Library}
}

@article{golzadeh2021ground,
  title={A ground-truth dataset and classification model for detecting bots in GitHub issue and PR comments},
  author={Golzadeh, Mehdi and Decan, Alexandre and Legay, Damien and Mens, Tom},
  journal={Journal of Systems and Software},
  volume={175},
  pages={110911},
  year={2021},
  publisher={Elsevier}
}

@inproceedings{franke2024exploratory,
  title={An exploratory mixed-methods study on general data protection regulation (gdpr) compliance in open-source software},
  author={Franke, Lucas and Liang, Huayu and Farzanehpour, Sahar and Brantly, Aaron and Davis, James C and Brown, Chris},
  booktitle={Proceedings of the 18th ACM/IEEE International Symposium on Empirical Software Engineering and Measurement},
  pages={325--336},
  year={2024}
}

@inproceedings{franke2024first,
  title={A First Look at the General Data Protection Regulation (GDPR) in Open-Source Software},
  author={Franke, Lucas and Liang, Huayu and Brantly, Aaron and Davis, James C and Brown, Chris},
  booktitle={Proceedings of the 2024 IEEE/ACM 46th International Conference on Software Engineering: Companion Proceedings},
  pages={268--269},
  year={2024}
}

@inproceedings{abdellatif2022bothunter,
  title={BotHunter: An approach to detect software bots in GitHub},
  author={Abdellatif, Ahmad and Wessel, Mairieli and Steinmacher, Igor and Gerosa, Marco A and Shihab, Emad},
  booktitle={Proceedings of the 19th International Conference on Mining Software Repositories},
  pages={6--17},
  year={2022}
}

@misc{unctaddata,
  author = {},
  title = {UN Trade and Development, Data Protection and Privacy Legislation Worldwide},
  howpublished = "\url{https://unctad.org/page/data-protection-and-privacy-legislation-worldwide}",
  year = {n.d.}, 
  note = "[Online; accessed 3-Apr-2026]"
}

@inproceedings{kapitsaki2024gdpr,
  title={GDPR indications in commits messages in GitHub repositories},
  author={Kapitsaki, Georgia and Papoutsoglou, Maria},
  booktitle={Proceedings of the 2024 IEEE/ACM 46th International Conference on Software Engineering: Companion Proceedings},
  pages={350--351},
  year={2024}
}

@misc{securiti,
  author = {Securiti AI},
  title = {CCPA vs GDPR},
  howpublished = "\url{https://securiti.ai/ccpa-vs-gdpr/}",
  year = {n.d.}, 
  note = "[Online; accessed 3-Apr-2026]"
}

@misc{txmatching968,
  author = {1338777555},
  title = {},
  howpublished = "\url{https://github.com/mild-blue/txmatching/issues/968}",
  year = {2022},
  note = "[Online; accessed 3-Apr-2026]"
}

@misc{ElasticPress3316,
  author = {1584610407},
  title = {},
  howpublished = "\url{https://github.com/10up/ElasticPress/issues/3316}",
  year = {2023},
  note = "[Online; accessed 3-Apr-2026]"
}

@misc{matomo-for-wordpress826,
  author = {1831381900},
  title = {},
  howpublished = "\url{https://github.com/matomo-org/matomo-for-wordpress/issues/826}",
  year = {2023},
  note = "[Online; accessed 3-Apr-2026]"
}

@misc{software-engineering6892,
  author = {976127507},
  title = {},
  howpublished = "\url{https://github.com/remote-job-boards/software-engineering/issues/6892}",
  year = {2021},
  note = "[Online; accessed 3-Apr-2026]"
}

@misc{openNDS62,
  author = {2267081316},
  title = {},
  howpublished = "\url{https://github.com/w3c/dpv/issues/140}",
  year = {2024},
  note = "[Online; accessed 3-Apr-2026]"
}

@misc{vertx-auth102,
  author = {188962803},
  title = {},
  howpublished = "\url{https://github.com/eclipse-vertx/vertx-auth/issues/102}",
  year = {2016},
  note = "[Online; accessed 3-Apr-2026]"
}

@misc{slackdump126,
  author = {1356890981},
  title = {},
  howpublished = "\url{https://github.com/rusq/slackdump/issues/126}",
  year = {2022},
  note = "[Online; accessed 3-Apr-2026]"
}

@inproceedings{tufano2024unveiling,
  title={Unveiling ChatGPT’s Usage in Open Source Projects: A Mining-based Study},
  author={Tufano, Rosalia and Mastropaolo, Antonio and Pepe, Federica and Dabi{\'c}, Ozren and Di Penta, Massimiliano and Bavota, Gabriele},
  booktitle={2024 IEEE/ACM 21st International Conference on Mining Software Repositories (MSR)},
  pages={571--583},
  year={2024},
  organization={IEEE}
}

@misc{nextcloudregistration1,
  author = {327781012},
  title = {},
  howpublished = "\url{https://github.com/nextcloud/registration/issues/140}",
  year = {2018},
  note = "[Online; accessed 3-Apr-2026]"
}

@misc{arx126,
  author = {251140333},
  title = {},
  howpublished = "\url{https://github.com/arx-deidentifier/arx/issues/126}",
  year = {2017},
  note = "[Online; accessed 3-Apr-2026]"
}

@misc{calypso22512,
  author = {297623638},
  title = {},
  howpublished = "\url{https://github.com/Automattic/wp-calypso/issues/22512}",
  year = {2018},
  note = "[Online; accessed 3-Apr-2026]"
}

@misc{sdk-sample-code-ios2,
  author = {313743203},
  title = {},
  howpublished = "\url{https://github.com/kshtgarg21/sdk-sample-code-ios/pull/2} (Pull request referenced in issue)",
  year = {2018},
  note = "[Online; accessed 3-Apr-2026]"
}

@misc{sdk-android153,
  author = {319506778},
  title = {},
  howpublished = "\url{https://github.com/schibsted/account-sdk-android/issues/153}",
  year = {2018},
  note = "[Online; accessed 3-Apr-2026]"
}

@misc{couchers2399,
  author = {1107664146},
  title = {},
  howpublished = "\url{https://github.com/Couchers-org/couchers/issues/2399}",
  year = {2022},
  note = "[Online; accessed 3-Apr-2026]"
}

@misc{dokuwiki2321,
  author = {315863357},
  title = {},
  howpublished = "\url{https://github.com/dokuwiki/dokuwiki/issues/2321}",
  year = {2018},
  note = "[Online; accessed 3-Apr-2026]"
}

@misc{GDPRDialog49,
  author = {328761675},
  title = {},
  howpublished = "\url{https://github.com/MFlisar/GDPRDialog/issues/49}",
  year = {2018},
  note = "[Online; accessed 3-Apr-2026]"
}

@misc{Kunena-Forum6032,
  author = {339454214},
  title = {},
  howpublished = "\url{https://github.com/Kunena/Kunena-Forum/issues/6032}",
  year = {2018},
  note = "[Online; accessed 3-Apr-2026]"
}

@misc{pubpub390,
  author = {453600833},
  title = {},
  howpublished = "\url{https://github.com/pubpub/pubpub/issues/390}",
  year = {2019},
  note = "[Online; accessed 3-Apr-2026]"
}

@misc{MultiChat70,
  author = {549727630},
  title = {},
  howpublished = "\url{https://github.com/MultiChat/Development/issues/70}",
  year = {2020},
  note = "[Online; accessed 3-Apr-2026]"
}

@misc{ceda-services-portal2,
  author = {675930545},
  title = {},
  howpublished = "\url{https://github.com/cedadev/ceda-services-portal/issues/2}",
  year = {2020},
  note = "[Online; accessed 3-Apr-2026]"
}

@misc{spectre402,
  author = {551721211},
  title = {},
  howpublished = "\url{https://github.com/dhowe/spectre/issues/402}",
  year = {2020},
  note = "[Online; accessed 3-Apr-2026]"
}

@misc{DMPonline_v427,
  author = {464312955},
  title = {},
  howpublished = "\url{https://github.com/DMPbelgium/DMPonline_v4/issues/27
}",
  year = {2019},
  note = "[Online; accessed 3-Apr-2026]"
}

@misc{django-newsfeed8,
  author = {698244572},
  title = {},
  howpublished = "\url{https://github.com/saadmk11/django-newsfeed/issues/8}",
  year = {2020},
  note = "[Online; accessed 3-Apr-2026]"
}

@misc{Open-Web-Analytics683,
  author = {708451536},
  title = {},
  howpublished = "\url{https://github.com/Open-Web-Analytics/Open-Web-Analytics/issues/683}",
  year = {2020},
  note = "[Online; accessed 3-Apr-2026]"
}

@misc{poingstudios81,
  author = {1689709220},
  title = {},
  howpublished = "\url{https://github.com/poingstudios/godot-admob-plugin/issues/81}",
  year = {2023},
  note = "[Online; accessed 3-Apr-2026]"
}

@misc{newrelic734,
  author = {841430967},
  title = {},
  howpublished = "\url{https://github.com/newrelic/opensource-website/issues/734}",
  year = {2021},
  note = "[Online; accessed 3-Apr-2026]"
}

@misc{gebruederheitz8,
  author = {1004140497},
  title = {},
  howpublished = "\url{https://github.com/gebruederheitz/consent-tools/issues/8}",
  year = {2021},
  note = "[Online; accessed 3-Apr-2026]"
}

@misc{re-cite35,
  author = {592152652},
  title = {},
  howpublished = "\url{https://github.com/MargaretKrutikova/re-cite/issues/35}",
  year = {2020},
  note = "[Online; accessed 3-Apr-2026]"
}

@misc{Adyen2403,
  author = {1970202037},
  title = {},
  howpublished = "\url{https://github.com/Adyen/adyen-web/issues/2403}",
  year = {2023},
  note = "[Online; accessed 3-Apr-2026]"
}

@misc{open-booking-api99,
  author = {412953164},
  title = {},
  howpublished = "\url{https://github.com/openactive/open-booking-api/issues/99}",
  year = {2019},
  note = "[Online; accessed 3-Apr-2026]"
}

@misc{city-grind13,
  author = {2050330189},
  title = {},
  howpublished = "\url{https://github.com/JJAMES24/city-grind/issues/13}",
  year = {2023},
  note = "[Online; accessed 3-Apr-2026]"
}

@misc{AspNetCoreDocs8174,
  author = {352206369},
  title = {},
  howpublished = "\url{https://github.com/dotnet/AspNetCore.Docs/issues/8174}",
  year = {2018},
  note = "[Online; accessed 3-Apr-2026]"
}

@misc{IdentityServer43841,
  author = {525028491},
  title = {},
  howpublished = "\url{https://github.com/DuendeArchive/IdentityServer4/issues/3841}",
  year = {2019},
  note = "[Online; accessed 3-Apr-2026]"
}

@misc{veronalabs38,
  author = {366396812},
  title = {},
  howpublished = "\url{https://github.com/veronalabs/wp-sms/issues/38}",
  year = {2018},
  note = "[Online; accessed 3-Apr-2026]"
}

@misc{uradotdesign24,
  author = {408814297},
  title = {},
  howpublished = "\url{https://github.com/uradotdesign/works/issues/24}",
  year = {2019},
  note = "[Online; accessed 3-Apr-2026]"
}

@misc{cubecart2338,
  author = {462094859},
  title = {},
  howpublished = "\url{https://github.com/cubecart/v6/issues/2338}",
  year = {2019},
  note = "[Online; accessed 3-Apr-2026]"
}

@inproceedings{guber2023empirical,
  title={Empirical Exploration of Open-Source Issues for Predicting Privacy Compliance},
  author={Guber, Jenny and Reinhartz-Berger, Iris and Litvak, Marina},
  booktitle={International Conference on Conceptual Modeling},
  pages={63--73},
  year={2023},
  organization={Springer}
}

@mastersthesis{datler2023intended,
  title={Intended Compliance: An Automated Analysis of GDPR-related GitHub Issues},
  author={Datler, Elias},
  year={2023},
  school={ETH Zurich}
}

@article{landis1977measurement,
  title={The measurement of observer agreement for categorical data},
  author={Landis, J Richard and Koch, Gary G},
  journal={biometrics},
  pages={159--174},
  year={1977},
  publisher={JSTOR}
}

@article{granata2024gdpr,
  title={GDPR compliance through standard security controls: An automated approach},
  author={Granata, Daniele and Mastroianni, Michele and Rak, Massimiliano and Cantiello, Pasquale and Salzillo, Giovanni},
  journal={Journal of High Speed Networks},
  number={Preprint},
  pages={1--28},
  year={2024},
  publisher={IOS Press}
}

@article{tahaei2022understanding,
  title={Understanding privacy-related advice on stack overflow},
  author={Tahaei, Mohammad and Li, Tianshi and Vaniea, Kami},
  journal={Proceedings on Privacy Enhancing Technologies},
  year={2022}
}

@article{sangaroonsilp2023taxonomy,
  title={A taxonomy for mining and classifying privacy requirements in issue reports},
  author={Sangaroonsilp, Pattaraporn and Dam, Hoa Khanh and Choetkiertikul, Morakot and Ragkhitwetsagul, Chaiyong and Ghose, Aditya},
  journal={Information and Software Technology},
  volume={157},
  pages={107162},
  year={2023},
  publisher={Elsevier}
}

@misc{nopCommerce7155,
  author = {2261182415},
  title = {},
  howpublished = "\url{https://github.com/nopSolutions/nopCommerce/issues/7155}",
  year = {2024},
  note = "[Online; accessed 3-Apr-2026]"
}

@misc{zitmall6,
  author = {2261614823},
  title = {},
  howpublished = "\url{https://github.com/dcsndevs/zitmall/issues/6}",
  year = {2024},
  note = "[Online; accessed 3-Apr-2026]"
}

@misc{bigcommerce222,
  author = {562358232},
  title = {},
  howpublished = "\url{https://github.com/bigcommerce/checkout-js/pull/222} (Pull request referenced in issue)",
  year = {2020},
  note = "[Online; accessed 3-Apr-2026]"
}

@misc{Human-Connection330,
  author = {426181359},
  title = {},
  howpublished = "\url{https://github.com/Human-Connection/Human-Connection/issues/330}",
  year = {2019},
  note = "[Online; accessed 3-Apr-2026]"
}

@misc{Artaxerxes85,
  author = {562022954},
  title = {},
  howpublished = "\url{https://github.com/evolution-events/Artaxerxes/issues/85}",
  year = {2020},
  note = "[Online; accessed 3-Apr-2026]"
}

@misc{matomo15574,
  author = {565958439},
  title = {},
  howpublished = "\url{https://github.com/matomo-org/matomo/issues/15574}",
  year = {2020},
  note = "[Online; accessed 3-Apr-2026]"
}

@misc{githubterms,
  author = {GitHub},
  title = {GitHub Terms of Service},
  howpublished = "\url{https://docs.github.com/en/site-policy/github-terms/github-terms-of-service#h-api-terms}",
  year = {2025},
  note = "[Online; accessed 3-Apr-2026]"
}

@article{hadar2018privacy,
  title={Privacy by designers: software developers’ privacy mindset},
  author={Hadar, Irit and Hasson, Tomer and Ayalon, Oshrat and Toch, Eran and Birnhack, Michael and Sherman, Sofia and Balissa, Arod},
  journal={Empirical Software Engineering},
  volume={23},
  pages={259--289},
  year={2018},
  publisher={Springer}
}

@article{determann2020california,
  title={The California privacy rights act of 2020: a broad and complex data processing regulation that applies to businesses worldwide},
  author={Determann, Lothar and Tam, Jonathan},
  journal={Journal of Data Protection \& Privacy},
  volume={4},
  number={1},
  pages={7--21},
  year={2020},
  publisher={Henry Stewart Publications}
}

@inproceedings{de2022ensuring,
  title={Ensuring privacy in the application of the Brazilian general data protection law (LGPD)},
  author={de Castro, Evandro Thalles Vale and Silva, Geovana RS and Canedo, Edna Dias},
  booktitle={Proceedings of the 37th ACM/SIGAPP Symposium on Applied Computing},
  pages={1228--1235},
  year={2022}
}

@inproceedings{tesfay2018privacyguide,
  title={PrivacyGuide: towards an implementation of the EU GDPR on internet privacy policy evaluation},
  author={Tesfay, Welderufael B and Hofmann, Peter and Nakamura, Toru and Kiyomoto, Shinsaku and Serna, Jetzabel},
  booktitle={Proceedings of the fourth ACM international workshop on security and privacy analytics},
  pages={15--21},
  year={2018}
}

@inproceedings{kounoudes2024right,
  title={i-Right: Identifying and Classifying GDPR User Rights in Fitness Tracker and Smart Home Privacy Policies},
  author={Kounoudes, Alexia Dini and Kapitsaki, Georgia M and Katakis, Ioannis},
  booktitle={International Conference on Web Information Systems Engineering},
  pages={243--254},
  year={2024},
  organization={Springer}
}

@misc{replication,
  author = {Kapitsaki, Georgia and Papoutsoglou, Maria and Treude, Christoph and Theophilou, Ioanna},
  title = {},
  howpublished = "\url{https://github.com/gkapi/github-privacy-law-issues-analysis}",
  year = {2025}
}

@inproceedings{ferreyra2025good,
  title={The Good, the Bad, and the (Un) Usable: a Rapid Literature Review on Privacy as Code},
  author={Ferreyra, Nicol{\'a}s E D{\'\i}az and Khelifi, Sirine and Arachchilage, Nalin and Scandariato, Riccardo},
  booktitle={2025 IEEE/ACM 18th International Conference on Cooperative and Human Aspects of Software Engineering (CHASE)},
  pages={173--178},
  year={2025},
  organization={IEEE}
}

@article{volini2019perspective,
  title={A perspective on technology education for law students},
  author={Volini, Anthony},
  journal={Santa Clara High Tech. LJ},
  volume={36},
  pages={33},
  year={2019},
  publisher={HeinOnline}
}

@article{kapitsaki2025evolution,
  title={Evolution of repositories and privacy laws: commit activities in the GDPR and CCPA era},
  author={Kapitsaki, Georgia M and Papoutsoglou, Maria},
  journal={Journal of Systems and Software},
  pages={112515},
  year={2025},
  publisher={Elsevier}
}

@misc{mailchimp-for-wordpress556,
  author = {321736859},
  title = {},
  howpublished = "\url{https://github.com/ibericode/mailchimp-for-wordpress/issues/556}",
  year = {2018},
  note = "[Online; accessed 3-Apr-2026]"
}

@book{cohen2013statistical,
  title={Statistical power analysis for the behavioral sciences},
  author={Cohen, Jacob},
  year={2013},
  publisher={routledge}
}

@inproceedings{madampe2025we,
  title={How Are We Doing With Using AI-Based Programming Assistants For Privacy-Related Code Generation? The Developers' Experience},
  author={Madampe, Kashumi and Grundy, John and Arachchilage, Nalin},
  booktitle={Proceedings of the 29th International Conference on Evaluation and Assessment in Software Engineering},
  pages={684--689},
  year={2025}
}

@article{hazra2017using,
  title={Using the confidence interval confidently},
  author={Hazra, Avijit},
  journal={Journal of thoracic disease},
  volume={9},
  number={10},
  pages={4125},
  year={2017}
}

@inproceedings{kapitsaki2023privacy,
  title={A privacy policies dataset in Greek in the GDPR era},
  author={Kapitsaki, Georgia and Papoutsoglou, Maria},
  booktitle={Proceedings of the 27th Pan-Hellenic Conference on Progress in Computing and Informatics},
  pages={199--205},
  year={2023}
}

@misc{kapitsaki2025github_privacy_commits,
  title        = {GitHub Data Privacy Commits from JSS 2025},
  author       = {Georgia Kapitsaki and Maria Papoutsoglou},
  howpublished = {Zenodo},
  year         = {2025},
  doi          = {10.5281/zenodo.15532947},
  url          = {https://doi.org/10.5281/zenodo.15532947}
}

@inproceedings{kapitsaki2024exploratory,
  title={An Exploratory Study on Soft Skills present in Software Positions in Cyprus: a quasi-Replication Study},
  author={Kapitsaki, Georgia and Chatzivasili, Loukas and Papoutsoglou, Maria and Galster, Matthias},
  booktitle={Proceedings of the 18th ACM/IEEE International Symposium on Empirical Software Engineering and Measurement},
  pages={200--211},
  year={2024}
}

@book{krippendorff2018content,
  title={Content analysis: An introduction to its methodology},
  author={Krippendorff, Klaus},
  year={2018},
  publisher={Sage publications}
}

@article{hennig2026whos,
  title={The whos, whats, and whys of issues related to personal data and data protection in open-source projects on GitHub},
  author={Hennig, Anne and Schulte, Lukas and Herbold, Steffen and Kulyk, Oksana and Mayer, Peter},
  journal={Empirical Software Engineering},
  volume={31},
  number={1},
  pages={1--51},
  year={2026},
  publisher={Springer}
}

@inproceedings{devlin2019bert,
  title={Bert: Pre-training of deep bidirectional transformers for language understanding},
  author={Devlin, Jacob and Chang, Ming-Wei and Lee, Kenton and Toutanova, Kristina},
  booktitle={Proceedings of the 2019 conference of the North American chapter of the association for computational linguistics: human language technologies, volume 1 (long and short papers)},
  pages={4171--4186},
  year={2019}
}

@article{zhang2025bert,
  title={Do BERT-Like Bidirectional Models Still Perform Better on Text Classification in the Era of LLMs?},
  author={Zhang, Junyan and Huang, Yiming and Liu, Shuliang and Gao, Yubo and Hu, Xuming},
  journal={arXiv preprint arXiv:2505.18215},
  year={2025}
}

@misc{icgcargo839,
  author = {515712760},
  title = {},
  howpublished = "\url{https://github.com/icgc-argo/platform-ui/issues/839}",
  year = {2019},
  note = "[Online; accessed 3-Apr-2026]"
}

@article{woolson1986sample,
  title={Sample size for case-control studies using Cochran's statistic},
  author={Woolson, Robert F and Bean, Judy A and Rojas, Patricio B},
  journal={Biometrics},
  pages={927--932},
  year={1986},
  publisher={JSTOR}
}

@incollection{chandra2019inductive,
  title={Inductive coding},
  author={Chandra, Yanto and Shang, Liang},
  booktitle={Qualitative research using R: A systematic approach},
  pages={91--106},
  year={2019},
  publisher={Springer}
}

@misc{microjam97,
  author = {339804399},
  title = {},
  howpublished = "\url{https://github.com/cpmpercussion/microjam/issues/97}",
  year = {2018},
  note = "[Online; accessed 3-Apr-2026]"
}

@inproceedings{ferreira2023rulekeeper,
  title={RuleKeeper: GDPR-aware personal data compliance for web frameworks},
  author={Ferreira, Mafalda and Brito, Tiago and Santos, Jos{\'e} Fragoso and Santos, Nuno},
  booktitle={2023 IEEE Symposium on Security and Privacy (SP)},
  pages={2817--2834},
  year={2023},
  organization={IEEE}
}

@misc{aptible16,
  author = {539746550},
  title = {},
  howpublished = "\url{https://github.com/aptible/www/pull/16/}",
  year = {2019},
  note = "[Online; accessed 3-Apr-2026]"
}

@misc{moral-core22,
  author = {4184241445},
  title = {},
  howpublished = "\url{https://github.com/richardkfm/moral-core/pull/22}",
  year = {2026},
  note = "[Online; accessed 11-May-2026]"
}

@misc{trello-api-skill,
  author = {4194250102},
  title = {},
  howpublished = "\url{https://github.com/DarkbyteAT/trello-api-skill/pull/4}",
  year = {2026},
  note = "[Online; accessed 11-May-2026]"
}

\end{document}